\documentclass{amsart}
\usepackage{amsmath,amssymb,amscd}
\usepackage{amsthm}
\usepackage{url}
\usepackage{color}
\usepackage{amssymb}
\usepackage{amsfonts}

\newcommand{\sgn}{\operatorname{sgn}}

\theoremstyle{theorem}
\newtheorem{thm}{Theorem}[section]

\theoremstyle{definition}

\theoremstyle{remark}

\begin{document}
\title[Berezin integral as a Riemann sum]{Berezin integral as a Riemann sum} \author{Thomas Scanlon}
 \address{Department of Mathematics \\
 University of Californa, Berkeley \\
 Evans Hall \\
 Berkeley, CA 94720-3840 \\
 USA}
 \email{scanlon@math.berkeley.edu}
\author{Roman Sverdlov}
\address{Department of Mathematics and Statistics \\
University of New Mexico \\
Science and Math Learning Center, 230 \\
311 Terrace NE MSC01 1115 \\
Albuquerque, NM 87131-0001 \\
USA}
\email{rsverdlov@unm.edu}

\begin{abstract}
Berezin integration of functions of anticommuting Grassmann variables
is usually seen as a formal operation, sometimes even defined via 
differentiation.  Using the formalism of geometric algebra and geometric
calculus in which the Grassmann numbers are endowed with a second 
associative product coming from a Clifford algebra structure, we 
show how Berezin integrals can be realized in the high dimensional limit
as integrals in the sense of geometric calculus. We then show how the concepts of spinors and superspace transform into this framework.
\end{abstract}

\maketitle

\section{Introduction}
Berezin integration of functions of noncommuting variables is usually defined 
formally.  The behavior of this operator is 
determined by general properties one wishes it to have.
For example, it should be  
translation invariant 
\begin{equation}
\int d \theta f(\theta + \eta)  = \int d \theta f(\theta)
\end{equation}
where $f$ is a function of the anticommuting variable $\theta$ and 
$\eta$ is an anticommuting constant. This in particular implies 
\begin{equation}
\int d \theta (\theta + \eta)  = \int d \theta \theta
\end{equation}
and therefore 
\begin{equation}
\int d \theta \eta  = 0
\end{equation}
which, in light of the fact that $\eta$ is a constant, is conventionally assumed to imply 
\begin{equation}
\int d \theta =0
\end{equation}
(although, if one is more careful, one would note that it only implies that this integral is annihilated by $\eta$). Since $\theta \wedge \theta =0$ a general analytic function is $f(\theta) = a + b \theta$ so since the integral of a constant is zero, the only way to avoid all integrals being zero is to make integral of $\theta$ non-zero. The standard normalization convention is
\begin{equation}
\int d \theta \theta = 1 \text{ .}
\end{equation}
It follows that the Berezin integral behaves like a differential 
operator.    In particular, for a Grassmann variable $\theta$ and complex numbers $a$ and $b$, one has 
\begin{equation}
\int d \theta (a + b \theta)   =  b
\end{equation}

This implies a strange change of variables formula for integrals of 
exponentials.  Since $\theta \wedge \theta = 0$, for any complex number $k$ we have the 
finite Taylor expansion 
\begin{equation}
e^{k \theta} = \sum_{n=0}^\infty \frac{1}{n!} k^n \theta^n = 
1 + k \theta
\end{equation}
so that 
\begin{equation}
\int d \theta e^{k\theta} = k
\end{equation}
rather than $\frac{1}{k}$ as one might expect. 

The differential nature of the Berezin integral is sometimes taken as 
its definition.  For 
a function $f(\theta)$ of the anticommuting variable $\theta$ we write 
$\frac{\partial f}{\partial \theta}$ for the right derivative
defined by the rule that 
\begin{equation}
f(\theta + \eta) = f(\theta) + \frac{\partial f}{\partial \theta} \eta + o(|\eta|)
\end{equation}
where $\eta$ is a small Grassmann-valued increment to $\theta$. Then the Berezin integral may be expressed as a derivative
\begin{equation}
\int d \theta f(\theta) = \frac{d f}{d \theta}
\end{equation}

Berezin integration plays a 
central role in theories of path integration in quantum field theory~\cite[Section 9.5]{PeSc}, the theories of supersymmetry~\cite[Part 2]{Dine} and even statistical mechanics \cite{Statistical1, Statistical2}, but
the warning appears throughout the literature that this operator is merely 
formal and should not (even, cannot) be regarded as a genuine integral.  

There have, however, been some attempts to relate Berezin integrals to what we are familiar with in calculus. Many of these are based on proving theorems within more familiar contexts, and then extrapolating them to Berezin integrals \cite{RefereeReference1,RefereeReference2,RefereeReference3,RefereeReference4}. This may leave one wondering whether Berezin integrals can be reduced to Riemann sum in an explicit way, without appeals to extrapolation. One attempt to do so was carried out by Jeffrey Rabin \cite{Rabin}. He proposed to model the Berezin integral as a contour integral. However, he ran into a few problems that he acknowledged in that paper. For one thing, the product of anticommuting numbers, despite being commuting, $\theta_3 (\theta_1 \theta_2) = (\theta_1 \theta_2) \theta_3$ is not real: $(\theta_1 \theta_2)^2 =0$. So, in order for the integral to be real, despite $\theta d \theta$ looking like a product, he takes the Berezin integral and the corresponding Riemann integral to be different:  the former is some type of projection of the latter. The other problem that he encountered is the one with normalization, where he merely postulated that the normalization is $1$. In this paper we propose to make those two steps more natural in the following way. We appeal to the idea of geometric calculus  (see~\cite{HS, LDG} for an introduction) that combines two distinct products (Clifford product and the exterior product) into a single space. In this framework, the Clifford product allows us to do the kind of projection that Rabin was looking for. In particular, the Clifford product is used between ``infinitesimal" and ``finite" parts of the integral, while the wedge product is used within the finite part. We then use the geometric calculus version of the divergence theorem \cite{HS, LDG} to prove the relation between the Berezin integral and the derivative. As far as normalization, we do not view it as convention but, instead, we view it as a consequence of a specific domain of integration and its measure. In particular, the integrals hitherto assumed $1$ will no longer be $1$ over rescaled domain. This, in particular, explains why the integral of $e^{k \theta}$ is $k$, despite the fact that scaling properties would seem to suggest otherwise (see Eq \ref{Scaling}). While the work of Rabin is focused on contour integration, this paper is instead focused on integrals over closed surfaces (Section \ref{SingleSurface}) and directed volume (Section \ref{DirectedVolume}). Nevertheless, as hinted earlier, the concept can be applied to the contour integration as well and, indeed, the second author of this paper wrote a separate paper that is focused on the contour integrals \cite{ContourSverdlov}. 

 Our comparison 
between Berezin integrals and these integrals of geometric calculus is made 
through the following theorem. We introduce the notation and conventions
we require in the body of this paper, but let us express the main theorem here.
Let $e_1, e_2, e_3, \ldots$ be a standard sequence of generators for the 
Grassmann numbers.

Our main theorem is as follows.

\begin{thm}
\label{mainthm}
If $f(\theta_1, \ldots, \theta_n)$ is an analytic function of the anticommuting variables $\theta_1, \ldots, \theta_n$ 
and for each natural number $D$, $X_D \subseteq 
(\bigoplus_{i=1}^D \mathbb{R} e_i)^n$ is a region of volume 
$\frac{1}{D}$ having a smooth boundary $\partial X_D$, then 

\begin{equation}
\label{mainidentity}
\int^{Ber} d \theta_1 \cdots d \theta_n f(\theta_1, \ldots, \theta_n) 
= \lim_{D \to \infty}  \int_{\partial X_D} d \theta_1 * \cdots * d \theta_n * f
= \lim_{D \to \infty}  \int_{\partial X_D} (d \theta_1 \wedge \cdots \wedge d \theta_n) *f  \nonumber \end{equation}
\begin{equation} = \lim_{D \to \infty} \int d_{\mu_D} \theta_1 * \cdots * d_{\mu_D} \theta_n* f = \lim_{D \rightarrow \infty} \int (d_{\mu_D} \theta_1 \wedge \cdots \wedge d_{\mu_D} \theta_n) * f \end{equation}
The integral on the left of the first line is the Berezin integral. The other two integrals on the first line are the 
directed integrals in the sense of geometric calculus (see Section \ref{SingleSurface}). In the integral on the middle of the first line the product between differentials is Clifford and for the integral at the right of the first line the product between differentials is anticommuting wedge product. The integrals on the second line are volume integrals where volume measure, $\mu_D$, has a direction (see Section \ref{DirectedVolume}) and, again, two different products are used in the infinitesimal parts in those two integrals. 
\end{thm}

As one can see from the above theorem, we are trying to compare and contrast different ways of introducing Riemann sums, instead of focusing on just one approach. Our two main approaches are surface integrals (Sections \ref{SingleSurface} -- \ref{highdimlimit}) and directed volume integral (Sec \ref{DirectedVolume}). In \ref{Misc1} and \ref{Misc2} we consider other approaches. In Section \ref{ComplexGrassmannNumbers} the Grassmann numbers are extended to include complex numbers as well as spinors, and it is shown how the expected symmetries can be accommodated in the context of our model. Then in Section \ref{SUSY} it is shown how superspace can be constructed within our model. 

One could separately use volume integrals (Sections \ref{ComplexGrassmannNumbers} and \ref{SUSYVolume}) and surface integrals (Sections \ref{SingleSurface} -- \ref{highdimlimit}). As it turns out, surface integration leads to the integration over a higher dimensional torus-like shape, while volume integration remains an integral over the full space. Of the two, volume integration is the one that respects the symmetry transformations and, in this sense, it is preferable. On the other hand, for volume integration we must attach a direction to volume elements which is a bit unusual, and this is what motivated us to introduce surface integrals since a directed area element is somewhat more common. Due to the relative advantages of each approach, we chose to consider both kinds of integrals. In any case, the actual superspace is the full vector space -- not the torus, and its definition (sec \ref{Orbit}) remains the same regardless of the integrals we choose. Our choice is whether to integrate over the full superspace (volume case) or a subset (surface case). 

Finally, in Section \ref{ExamplesNonlinear} we briefly describe how supermathematics can be extended to include non-analytic functions within the framework of the models we introduce and how some of those non-analytic functions can be integrated. Then, in Section \ref{PhysicsNonlinear} we explore how non-analytic functions can be used in some of the physics models appearing in some of the second authors ongoing work.

\section{Some background information} \label{Notation}

\subsection{Notation and conventions}

In this section we establish our notation and conventions.  The reader may wish to 
consult~\cite{Rogers, BerezinBook} for an introduction to the theory of supermanifolds 
and~\cite{LDG,Brom,HS} for introductions to the theory 
of geometric algebra.   As the notation is not uniform in the literature, we must
make some choices.

For a natural number $D$, let 

\begin{equation}
V_D  =  \bigoplus_{i=1}^D \mathbb{R} e^i
\end{equation}

where $e^i$ are orthonormal. We will work with variables $\theta$ ranging over $V_D$.  Such variable may be 
decomposed with respect to the standard basis of $V_D$ as 

\begin{equation}
\theta = \sum_{i=1}^D x^i e^i
\end{equation}

where the variables $x^i$ are real valued.   Thus, if we have a function 
$f:V_D \to V_D$ which we may see as a function of the variable $\theta$, we
may also regard $f$ as a function of the $D$ variables $x_1, \ldots, x_D$. 

From $V_D$ we may form the Grassmann algebra $\mathcal{G}_D$ as the exterior
algebra of $V_D$:

\begin{equation}
\mathcal{G}_D = \Lambda^\ast (V_D)
\end{equation}

There is a second associative product on $\mathcal{G}_D$ coming from the 
Clifford algebra $C(V_D, \langle ~ , ~ \rangle)$ generated by $V_D$ with the standard inner product defined by 
\begin{equation}
\langle e^i,  e^j \rangle =  \delta_{i,j}
\end{equation}
There is a standard isomorphism of vector spaces between $\mathcal{G}_D = \Lambda^\ast(V_D)$ and
$C(V_D, \langle ~, ~ \rangle)$ from which by transport of structure we may 
regard $\mathcal{G}_D$ as having the Clifford product, $*$, as well as the usual 
wedge product.  These products are related.  

For $a_1, \ldots, a_n \in V_D$, we have 

\begin{equation}
a_1 \wedge  \cdots \wedge a_n  = \frac{1}{n!} \sum_{\sigma \in \operatorname{Sym}(n)}
\operatorname{sgn}(\sigma) a_{\sigma(1)}* \cdots* a_{\sigma(n)}
\end{equation}
Similarly, we can define a symmetric product as 
\begin{equation}
\langle a_1   \cdots  a_n  \rangle = \frac{1}{n!} \sum_{\sigma \in \operatorname{Sym}(n)}
 a_{\sigma(1)} * \cdots * a_{\sigma(n)}
\end{equation}
which, in particular, means

\begin{equation}
\langle a, b \rangle = \frac{1}{2}(a*b + b*a)
\end{equation}

Combining these identities, we have 

\begin{equation}
a*b = a \wedge b + \langle a, b \rangle
\end{equation}

For $D \leq D'$ there is a natural inclusion $V_D \hookrightarrow V_{D'}$
which induces an inclusion $\mathcal{G}_D \hookrightarrow \mathcal{G}_{D'}$
which is compatible with the Grassmann, Clifford and inner products. While we 
do not specifically use it ourselves, the direct 
limit $\mathcal{G} := \varinjlim_{D \to 
\infty} \mathcal{G}_D$ may be seen as the algebra of all Grassmann numbers. 

\subsection{Notational remarks regarding product signs} \label{ProductSigns}

 It is important to note that the notation of the Clifford product is different in different papers, including the papers by the second author \cite{ContourSverdlov,Sv2,Sv1}. This is due to the fact that, within the conventional context of Berezin integration \cite{PeSc, Dine,Drees}, there is only one type of product: the anticommuting exterior product, which is written as $\theta_1 \theta_2$ without wedge. On the other hand, in the context of geometric calculus ~\cite{HS, LDG}, the notation $\theta_1 \theta_2$ is reserved for the Clifford product. In some of the earlier \url{arXiv} versions of the current paper (including, for example, version 5 -- see \cite{Version5}), the second author uses the dot-product, $\theta_1 \cdot \theta_2$, for the Clifford product; but this notation is used for the commuting inner product in geometric calculus ~\cite{HS, LDG} while said inner product has been denoted by $\langle \theta_1 \vert \theta_2 \rangle$ in some of those earlier versions of this paper (including \cite{Version5}). In light of these observations, it seems most convenient to use $\theta_1 * \theta_2$ for Clifford product; unfortunately, in the second author's earlier works on the subject, including \cite{Version5}, $\theta_1 * \theta_2$ appears as a notation for a general product -- which can be either Clifford or wedge. But now that we are made aware of those other papers, we propose to change notation and use $\theta_1 * \theta_2$ strictly for Clifford product. On the other hand, for the situation where a product can be either the Clifford product or wedge we will write $\theta_1 \theta_2$. It should be acknowledged though that $\theta_1 \theta_2$ has been used in other ways, as described earlier. 

\subsection{Notation for variables} \label{Var}

In geometric calculus usually the letter $x$ is used for a variable; however, instead of using $x$ we will use $\theta$. The reason for this is that we are proposing a model of the Berezin integral and $\theta$ is normally used as a variable in that integral. Additionally, in physics $x$ is used for something else: namely, a \emph{commuting} variable of dimension $4$. In fact, in the case of supersymmetry,  both $x$ and $\theta$ are used as separate variables. Since we do include supersymmetry (see Section \ref{SUSY}) this is an additional reason against using $x$ for the anticommuting variable. 

\subsection{Complexification} \label{Var}

We will work with complexifications of real vector spaces. For us, if $W$ is a real vector space, $W_{\mathbb{C}} = \mathbb{C} \otimes_{\mathbb{R}} W$ denotes the complexification of $W$. 

\subsection{Flatness of space} \label{SpaceIsFlat}

Our work is based on Theorem \ref{KeyTheorem} which, in spirit, is a divergence theorem. Even though the volume of the region is $1/D$ where $D \rightarrow \infty$, this volume is \emph{not} infinitesimal when $D$ is finite. In light of the curvature of space, the divergence theorem only works when volume \emph{is} infinitesimal for fixed, finite, $D$. Since this is not true, we have to, instead, assume that the space over which $\theta$ ranges is flat. Indeed, the Proof of Theorem \ref{KeyTheorem} is based on this assumption. For example, the definition of $I$ that is used in its proof would be ambiguous if the space were curved. Therefore, everything else that we write here, being based on this theorem, is also, by default, based on the assumption that space over which $\theta$ ranges is flat. This assumption is implicit from this point on. 

\subsection{Some subtle differences between variable vectors and basis vectors} \label{VariablesConstants}

It is important to stress that, while $\theta_i$ are variables (in particular, general vectors), $e^i$ are constants (in particular, unit vectors with pre-assigned directions). This, in particular, means that we can only integrate with respect to $\theta_i$-s, but not with respect to $e_i$-s.  This is an important contrast to the conventional superanalysis where $\theta_i$ were used both as variables and as basis vectors at the same time.  On the other hand, in multivariable calculus, there is a clear distinction between vector variables, $\vec{v}_1, \vec{v}_2, \cdots$ and basis vectors, $\hat{e}_1, \hat{e}_2, \cdots$. One of the main goals of this paper is to bring superanalysis closer to multivariable calculus, and this includes drawing a fundamental distinction between $\theta_k$ and $e_k$. In multivariable calculus there is, indeed, such a thing as change of variables: for example, one can change variables between $\vec{v}_1, \vec{v}_2, \cdots$ and variables $\vec{u}_1, \vec{u}_2, \cdots$; but the letters $e$ are never used for variables, since they have already been reserved for basis vectors. Similarly, in our case, we can change variables from $\theta_1, \theta_2, \cdots$ to $\theta_1^{\prime}, \theta_2^{\prime}, \cdots$, or we can change them from $\eta_1, \eta_2, \cdots$ to $\xi_1, \xi_2, \cdots$; but we would never use $e_i$ to denote them since this notation is reserved for constant unit vectors. 

This difference between conventional superanalysis (where $\theta$-s and $e$-s are put on the same level) and our version of it (where they are not) is connected to the other areas we are proposing to reinterpret. In conventional superanalysis, $\theta_k$ is not viewed as an element of a set but, instead, it is viewed as a symbol (this, in turn, is connected to the fact that the Berezin integral is viewed as a symbolic operation as opposed to as a limit of a Riemann sum). Consequently, a homogeneous polynomial is identified with an $n$-tuple, 
\begin{equation} (a_1, \cdots, a_n) = a_1 \theta_1 + \cdots + a_n \theta_n \end{equation}
and, within this context, it logically follows that $\theta_1, \cdots, \theta_n$ are basis vectors. On the other hand, in the approach we are proposing, we would like to view $\theta$ as a literal element of a set -- namely, a vector -- so that its values are distinguishable from each other. Consequently, 
\begin{equation} f_{a_1 \cdots a_n} (\theta_1, \cdots, \theta_n) = a_1 \theta_1 + \cdots + a_n \theta_n \end{equation}
 is no longer an $n$-tuple but, instead, it is a literal function on a literal set. Of course, we might say that the set of linear functions could be generated by the basis functions $\{f_1, f_2, \cdots \}$, 
\begin{equation} f_i (\theta_1, \cdots, \theta_n) = \theta_i \end{equation}
but $f_i$ is distinct from $\theta_i$; in particular, $f_1$ is a single function, while $\theta_1$ is a variable that can take different values distinguishable from each other. In order to define the set of values $\theta_1$ can take, we need to define a space $V_D$ which, in turn, is generated by basis vectors $e_1, \cdots, e_D$; as any basis vectors, they are \emph{not} subject to change, in sharp contrast to $\theta_k$. 

Another source of this difference can be seen in the way we treat the expression
\begin{equation} \theta = X_1 e_1 + \cdots + X_D e_D \label{ThetaAndX} \end{equation}
where $e_1, \cdots, e_D$ are anticommuting while $X_1, \cdots, X_D$ are commuting. Conventionally, one would be inclined to assume that $e_1, \cdots, e_D$ are variables while $X_1, \cdots, X_D$ are constants; we propose to do the opposite: we view $X_1, \cdots, X_D$ as variables and $e_1, \cdots, e_D$ as constants. This difference in views is due to the fact that, if $X_1, \cdots, X_D$ are to be variables, one would expect to express the integral over anticommuting variable $\theta$ in terms of the integrals over commuting variables $X_1, \cdots, X_D$ which, conventionally, is unthinkable. But, in the context of this paper, our goal is to be able to do just that. In particular, we define the directed volume  $d_{\mu} \theta$, as 
\begin{equation} d_{\mu} \theta = \frac{X_1 e_1 + \cdots + X_D e_D}{\sqrt{X_1^2 + \cdots + X_D^2}} \mu \Big(\sqrt{X_1^2 + \cdots + X_D^2} \Big) dX_1 \cdots dX_D \end{equation}
and then express the Berezin integral as 
\begin{equation} \int f (\theta) *d_{\mu} \theta = \label{CommutingIntegral}\end{equation}
\begin{equation} = \int f (x_1 e_1 + \cdots + x_D e_D) * \frac{X_1 e_1 + \cdots + X_D e_D}{\sqrt{X_1^2 + \cdots + X_D^2}} \mu \Big(\sqrt{X_1^2 + \cdots + X_D^2} \Big) dX_1 \cdots dX_D \nonumber \end{equation}
This allows for $X_1, \cdots, X_D$ to take upon the role of variables, which would allow for $e_1, \cdots, e_D$ to be unambiguously constants. In particular, in contrast to conventional superanalysis, we can integrate over $X_1, \cdots, X_D$, but we can \emph{not} integrate over $e_1, \cdots, e_D$.

As stated before, we \emph{do} have such a thing as change in variables where both sets of variables are anticommuting. For example, we can express the value of a Grassmann variable $\xi$, in terms of different bases, $\{e_1, \cdots, e_D \}$ and $\{e_1^{\prime}, \cdots, e_D^{\prime} \}$
\begin{equation} \xi = \sum_{k=1}^D X_k e_k = \sum_{k=1}^D X^{\prime}_k e^{\prime}_k \; , \; X^{\prime}_k = \sum_{l=1}^D M_{kl} X_l \; , \; e^{\prime}_k = \sum_{l=1}^D (M^{-1})_{kl} e_l \end{equation}
or we can define a new variables, $\theta_1^{\prime}, \cdots, \theta_n^{\prime}$, as a linear combinations of the old ones, $\theta_1, \cdots, \theta_n$: 
\begin{equation} \theta^{\prime}_k = \sum_{l=1}^n A_{kl} \theta_l \Longleftrightarrow \sum_{j=1}^D X^{\prime}_{kj} e_j = \sum_{l=1}^n \bigg(A_{kl} \sum_{j=1}^D X_{lj} e_j \bigg) \Longleftrightarrow  X_{kj}^{\prime} = \sum_{l=1}^n A_{kl} X_{lj} \end{equation}
This, however, does not change what we said a bit earlier. In the above expressions, $\{e_k \}$ and $\{e_k^{\prime} \}$ are anticommuting constants, $\{X_{kl} \}$ and $\{ X^{\prime}_{kl} \}$ are commuting variables, while $\xi$, $\{\theta_k \}$ and $\{ \theta^{\prime}_k \}$ are anticommuting variables. As a rule, the letter $e$ is reserved for anticommuting constants; letters $\theta$, $\eta$ and $\xi$ are reserved for anticommuting variables; the last letters of the latin alphabet refer to commuting variables, and the first letters in latin alphabet refer to commuting constants.

\section{Integration} \label{Integrals} 

In this section we consider various integration theories and then show how
to interpret the Berezin integral as a genuine geometric integral.

\subsection{Single integrals over the closed surface}
\label{SingleSurface} 

In the theory of directed integrals in the sense of geometric calculus (see~\cite[Chapter 7]{LDG} 
or~\cite[Chapter 4]{Brom}), we may make sense of integrals of $\mathcal{G}$-valued 
functions on manifolds embedded in $V_D$.  For us, the most important result in this theory 
is its version of the divergence theorem (see~\cite[Equation 6.150]{Brom}).    

We take $M \subseteq V_D$ to be an open region with smooth boundary 
$\partial M$ and $f:M \to C(V_D, \langle , \rangle)$ a smooth 
function.  We define 

\begin{equation}
\nabla \cdot f = \sum_{k=1}^D e^k \frac{\partial f}{\partial x^k}
\end{equation}
 
We then have the following theorem:
\begin{thm} \label{KeyTheorem}
 If we assume that the space is flat, 
\begin{equation}
\label{divthm}
\int_M \nabla \cdot f |dX| = \int_{\partial M} n* f |dS|
\end{equation}
\end{thm} 
where $n$ is a unit normal vector (where its norm, which happens to be $1$, is defined with respect to the Clifford product) that is normal to $\partial M$ pointing outward and $nf$ is the Clifford product between that vector and $f$.
\begin{proof}
Following \cite{Book}, define $I$ to be 
\begin{equation} I = e_1 \wedge \cdots \wedge e_d \end{equation}
and denote the dimensionality of a given element by the index at the bottom. For example, an area, being $d-1$ dimensional object, will be $S_{d-1}$. By using $I$ we can establish a correspondence of the type
\begin{equation} G_k = (-1)^{\sum_{j=d-k}^{d-1} j} I*G_{d-k} \end{equation}
We also borrow the notation from ~\cite{HS, LDG} that two matching dots -- one above the derivative and the other one above one of the functions -- indicates that the derivative is acting only on that function and on nothing else (for example, $\dot{\nabla} \dot{f} g$ is equal to $g \nabla f$ and the product rule is $\nabla (fg) = \dot{\nabla} \dot{f} g + \dot{\nabla} f \dot{g}$). With this notation, we can perform the following calculation: 
\begin{equation} \int_{\partial V} dS_1 * f_1 = (-1)^{d-1} \int_{\partial V} (dS_{d-1} *I)*f_1 =(-1)^{d-1} \int_{\partial V} dS_{d-1} * (I*f_1)  \nonumber \end{equation}
\begin{equation} = - \int_{\partial V} dS_{d-1} * f_{d-1} =-  \int_V \dot{\nabla} *dV_d * \dot{f}_{d-1} = (-1)^{d-1} \int_V \dot{\nabla} *dV_d  * (I * \dot{f}_1)  \nonumber \end{equation}
\begin{equation} = (-1)^{d-1} \int_V \dot{\nabla} (dV_d *I)* \dot{f_1} = \int_V \dot{\nabla} *dV_0 *\dot{f}_1 = \int_V \dot{\nabla} * \dot{f}_1 dV_0  \end{equation} 
where on the fourth equal sign we used Eq 6.163 of \cite{Book}, in the last equal sign we used the fact that $dV_0$ is a scalar and, therefore, commutes with everything. By noticing that 
\begin{equation} dS_1 = n \vert dS \vert \; , \; dV_0 = \vert dV \vert \end{equation}
the statement equating the left hand side with the right hand side of the above calculation becomes 
\begin{equation} \int_{\partial V} (n \vert dS \vert) * f_1 = \int_V \dot{\nabla} * \dot{f}_1 \vert dV \vert \end{equation} 
Since $\vert dS \vert$ is a scalar and, therefore, commutes with everything, we can rewrite it as
\begin{equation} \int_{\partial V} (n*f_1) \vert dS \vert  = \int_V \dot{\nabla} * \dot{f}_1 \vert dV \vert \end{equation}
which completes the proof. 

\end{proof}

{\bf Note:} The statement of the above theorem is, roughly speaking, an equivalent of Eq 6.163 of \cite{Book}. We have purposely utilized that equation in our proof in order to emphasize the connection between the two statements. Thus, a reader that is used to the presentation given in \cite{Book} can look at the above proof as a type of bridge between the notation used in \cite{Book} and the notation we will be using. 

 While the statement of the above theorem is written in the notation of geometric calculus, for our purposes, we would like to denote the variables of integration by $d \theta$. Notably, the integral on the left hand side is a volume integral while the integral on the right hand side is a surface integral. Thus, $d \theta$ changes its meaning depending on whether the integral is being taken over the surface or over the volume. In neither case is $d \theta$ a scalar. If the integral is taken over the surface, then $d \theta$ is an infinitesimal $V_D$-valued element, and if it is being taken over a volume, then it is infinitesimal $V_D^D$-valued element. However, $\vert d \theta \vert$ is an infinitesimal scalar: in one case it is a scalar area element, in the other case it is a scalar volume element. With this new notation, we rewrite the above equation as 
\begin{equation}
\label{divthm2}
\int_M \nabla \cdot f |d \theta| = \int_{\partial M} n* f |d \theta|
\end{equation}

 For such a function $f$ we define the \emph{average directional derivative} by

\begin{equation}
\bigg\langle \frac{d f}{d \theta} \bigg\rangle = \frac{1}{D} \sum_{i=1}^D e^i* \frac{\partial f}{\partial x^k} = \frac{1}{D} \nabla \cdot f
\end{equation}

From the definition we have 

\begin{equation}
\bigg\langle \frac{d \theta}{d \theta} \bigg\rangle = \frac{1}{D} \sum_{i=1}^D e^i * \frac{\partial \theta}{\partial x^i} = \frac{1}{D} \sum_{i=1}^D e^i * e^i = \frac{1}{D} D = 1
\end{equation}

Let us define 

\begin{equation}
\bigg\langle \frac{d f}{d \theta} \bigg\rangle_M = \frac{1}{\operatorname{vol}(M)} \int_M |d \theta| \bigg\langle \frac{d f}{d \theta} \bigg\rangle  \text{ ,}
\end{equation}

where $d \theta$ is $D$-volume element with values in $V_D^D$ and $\vert d \theta \vert$ is a real-valued volume element defined in terms of a norm of $d \theta$ where norm is defined in terms of Clifford product. Then we obtain

\begin{equation}
\label{avgint}
\int_{\partial M} d \theta * f(\theta) = \int_M |d \theta| \nabla \cdot f = \int_M |d \theta| D \bigg \langle \frac{d f}{d \theta} \bigg \rangle 
= \end{equation}
\begin{equation} = D \operatorname{vol}(M) \frac{1}{\operatorname{vol}(M)} \int_{M} |d \theta| \bigg \langle \frac{d f}{d \theta} \bigg \rangle = D \operatorname{vol}(M)
\bigg \langle \frac{d f}{d \theta} \bigg \rangle_M \nonumber
\end{equation}
where $d \theta$ changes its meaning depending on the domain of integration: in case of the surface integral over $\partial M$ it refers to a $V_D$-valued area element $ndS$, whereas in case of integration over $M$ it refers to $V_D^D$-valued volume element. 
If we take $\operatorname{vol}(M) = \frac{1}{D}$, then Equation~\ref{avgint} reduces to 

\begin{equation}
\label{nearlyBer}
\int_{\partial M} d \theta * f(\theta) = \bigg \langle \frac{d f}{d \theta} \bigg \rangle_M
\end{equation}

For analytic functions $f$, the expression $\bigg \langle \frac{d f}{d \theta} \bigg \rangle_M$ matches the Berezin integral of $f$.  

With this we complete the proof of Theorem~\ref{mainthm}
in the case where $n = 1$ and $f$ is analytic.

It is noteworthy that the condition $\operatorname{vol}(M) = \frac{1}{D}$ is the reason why the usual reparametrization properties do not apply: the scaled region $kM$ no longer meets that condition. Thus, we get 
\begin{equation}
\int_{\partial M} e^{k \theta} * d \theta = \frac{1}{k^{D-1}} \int_{\partial (kM)} e^{\theta} * d \theta = \frac{1}{k^{D-1}} k^{D} = k \label{Scaling} \end{equation}
The coefficient $\frac{1}{k^{D-1}}$ is what we would expect for a 
surface integral, but the coefficient $k^D$ is due to the fact that $kM$ does not meet the volume condition, so instead of $1$ we have $k^D$, and as a result the region that does meet the volume condition produce an integral of $k$, as expected. 

\subsection{Sign convention for multiple integrals}
\label{sect:signConvention}

Before we proceed with multiple integrals, it is important to make a note of the sign convention we will be using. Most books, including~\cite{PeSc}, use the sign convention 
\begin{equation}
\int d \theta_1 d \theta_2 \theta_1 \theta_2 = +1 \end{equation}
\begin{equation}
\int d \theta_1 d \theta_2 \theta_2 \theta_1 = -1 \end{equation}
(where we have skipped $*$ and $\wedge$ because the books in question are not using it). On the other hand, the book~\cite[page 20]{Drees} uses a different convention; namely, 
\begin{equation}
\int d \theta_1 d \theta_2 \theta_1 \theta_2 = -1 \end{equation}
\begin{equation}
\int d \theta_1 d \theta_2 \theta_2 \theta_1 = +1 \end{equation}
Even though the former convention is more commonly used, we regard the latter convention as more logical:
\begin{equation}
\int d \theta_1 d \theta_2 \theta_2 \theta_1 = \int \bigg(d \theta_1 \bigg(\int d \theta_2 \theta_2 \bigg) \theta_1 \bigg) = \int (d \theta_1 1 \theta_1) = \int d \theta_1 \theta_1 = 1 \end{equation}
and, for that reason, in this paper we will stick with the latter sign convention. 

\subsection{Equivalence between different types of multiple integrals over the closed surface}
\label{sect:multiple}

As one sees in Equation~\ref{mainidentity} multiple integrals come 
in two different forms, one in which the product between the 
differentials is the Clifford product and a second in which the 
product between the differentials is the usual wedge product. 
On the one hand, from the point of view of iterated integrals, 
the Clifford product is more appealing and, on the other hand, 
from the point of view of area elements the Grassmann product
is more appealing.  The main result of this section is that for
analytic integrands, it is not necessary for us to choose between
these two forms since the integrals are equal.  

% Rather than having to choose between those two
% expressions, let us prove that they are actually equivalent. 

As in Section~\ref{SingleSurface}, we take $M \subseteq V_D$ 
an open region with finite volume and a smooth 
boundary $\partial M$.

\begin{thm} \label{EitherProductTheorem}
\label{eq-different-integrals}
Let $f(\theta_1,\ldots,\theta_n)$ be an analytic function
of the noncommuting variables $\theta_1, \ldots, \theta_n$ we have the equalities

\begin{equation} \int_{ \partial(M)^n } (d \theta_1 \wedge \ldots \wedge  d \theta_n)* f(\theta_1, \ldots, \theta_n) = \int_{\partial(M)^n} d \theta_1 * \ldots *d  \theta_n* f(\theta_1, \ldots, \theta_n)   \nonumber \end{equation}
\begin{equation} =  \int_{\partial(M)} \bigg( d \theta_1 * \int_{\partial(M)} \bigg( d \theta_2 * \bigg( \cdots * \int_{\partial(M)} 
d \theta_n * f (\theta_1, \ldots, \theta_n) \bigg) \cdots \bigg) \bigg) \end{equation} \end{thm}

\begin{proof}

We start by computing the integrals with respect to Clifford products of 
differentials. 

Before doing so, we need to 
extend our notation for average 
directional derivatives to the 
situation of functions of several
variables.  If $g(\theta_1,
\ldots,\theta_n)$ is a 
function of $n$ noncommuting 
variables and $1 \leq k 
\leq n$, then we write 
$\bigg \langle \partial_{\theta_k} 
g \bigg \rangle$ for the average
directional derivative of $g$ 
regarded as a function of 
$\theta_k$ alone.

Consider the case of $F(\theta_1,\theta_2)$ a function 
of two variables.

Note that 
\begin{equation} \int_{\partial M \times \partial M} d \theta_1* d \theta_2 *F(\theta_1, \theta_2) = \int_{\partial M} \bigg(d \theta_1 *\int_{\partial M} d \theta_2 * F(\theta_1, \theta_2) \bigg)  = 
\operatorname{vol}(M) D \bigg\langle \partial_{\theta_1} \int_{\partial M} d \theta_2 * F(\theta_1, \theta_2) \bigg\rangle_M  \nonumber \end{equation}
\begin{equation} = (\operatorname{vol}(M) D)^2 \bigg\langle \partial_{\theta_1} \bigg\langle \partial_{\theta_2} F(\theta_1, \theta_2) \bigg\rangle_M \bigg\rangle_M = (\operatorname{vol}(M) D)^2 \bigg\langle \partial_{\theta_1} \partial_{\theta_2} F(\theta_1, \theta_2) \bigg\rangle_{M \times M} \end{equation}
By proceeding in the same way $n$ times for $F(\theta_1, \ldots,  \theta_n)$
a function of $n$ variables, we obtain 
\begin{equation} \int d \theta_1 * \ldots *d \theta_n * F(\theta_1, \ldots, \theta_n) = (\operatorname{vol}(M) D)^n \langle \partial_{\theta_1} \ldots \partial_{\theta_n} F(\theta_1, \ldots, \theta_n) \rangle_{M^n} \label{DotDerivative} \end{equation}

We shall now compare this calculation the result of computing the integral with 
respect to wedge products of the differentials. 

Recall our formula for computing wedge products from Clifford products.
\begin{equation} a_1 \wedge \ldots \wedge a_n = \frac{1}{n!} \sum_{\sigma} \operatorname{sgn}(\sigma) a_{\sigma(1)}* \ldots *a_{\sigma(n)} \label{Recall} \end{equation}
From this we obtain the following identity.
\begin{equation} \int (d \theta_1 \wedge \ldots \wedge d \theta_n) * F(\theta_1, \ldots, \theta_n) = \frac{(\operatorname{vol}(M) D)^n}{n!} \sum_{\sigma} \bigg( \operatorname{sgn}(\sigma) \int d \theta_{\sigma(1)}* \ldots * d \theta_{\sigma(n)} * F(\theta_1, \ldots, \theta_n) \bigg)   \nonumber  \end{equation}
\begin{equation} = \frac{(\operatorname{vol}(M) D)^n}{n!} \sum_{\sigma} \big( \operatorname{sgn}(\sigma)  \langle \partial_{\theta_{\sigma (1)}}  \ldots  \partial_{\theta_{\sigma (n)}} F \rangle_{M^n} \big) \label{WedgeDerivatives} \end{equation}

Thus, from Equations~\ref{DotDerivative} and~\ref{WedgeDerivatives}, we
see that to establish this theorem, we need to show that the two kinds
of derivative expressions are equal.  That is, we must prove the 
following identity.

\begin{equation}  \langle \partial_{\theta_1} \ldots \partial_{\theta_n} F(\theta_1, \ldots, \theta_n) \rangle_{M^n} = \frac{1}{n!} \sum_{\sigma} \operatorname{sgn}(\sigma) \langle \partial_{\theta_{\sigma(1)}} \ldots \partial_{\theta_{\sigma(n)}} F(\theta_1, \ldots, \theta_n) \rangle_{M^n}  \label{Equivalence1} \end{equation}

We proceed now to show that Equation~\ref{Equivalence1} holds.  

To do so we need to make use of the fact that $F$ is analytic, that is, 
it may be expressed as a power series in $\wedge$-monomials of the 
the variables $\theta_1, \ldots, \theta_n$.  Since these variables 
anticommute, for each $\theta_k$ and integer $p \geq 2$ we have 
$\theta_k^p = 0$ if the power is defined in terms of wedge-product.  Thus, to say that $F$ is analytic is to say we may 
write 
\begin{equation}
F(\theta_1, \ldots, \theta_n) = \sum_{l=0}^n \; \; \;  \sum_{1 \leq j_1 < \cdots < j_\ell \leq n}
c_{j_1, \ldots, j_\ell} \theta_{j_1} \wedge \cdots \wedge \theta_{j_\ell}
\nonumber
\end{equation}

for suitable constants $c_{j_1, \ldots, j_\ell}$.

Since the expressions on each side of Equation~\ref{Equivalence1} are linear,
it suffices to consider the case that $F$ is a basic monomial of the form 
$\theta_{j_1} \wedge \cdots \wedge \theta_{j_\ell}$ with $1 \leq j_1 < 
\cdots < j_\ell \leq n$.  Note that if $\ell < n$, then for some $k$ the variable 
$\theta_k$ does not appear in $F$ but on each side of Equation~\ref{Equivalence1}
we apply the operator $\partial_k$.  Thus, in this case, Equation~\ref{Equivalence1}
reverts to the tautology $0 = 0$.

Otherwise, $F$ is fully antisymmetric in the 
sense that for any permutation $\sigma$ of $\{ 1, \ldots, n \}$ we have 

\begin{equation}
F(\theta_{\sigma(1)}, \ldots, \theta_{\sigma(n)}) = \sgn(\sigma) F(\theta_1,
\ldots,\theta_n)
\nonumber
\end{equation}

provided that $F$ is a monomial. Since in Equation~\ref{Equivalence1} 
we are averaging on each side of the equality and because each of the $\theta_k$ is averaged over the same region, we 
may freely permute them.  Thus, we obtain for any permutation $\sigma$ 
of $\{1, \ldots, n \}$ the following equality.

\begin{equation} \langle \partial_{\theta_{\sigma(1)}} \ldots \partial_{\theta_{\sigma(n)}} F(\theta_1, \ldots, \theta_n)\rangle_{M^n} = \langle \partial_{\theta_1} \ldots \partial_{\theta_n} F(\theta_{\sigma^{-1} (1)}, \ldots, \theta_{\sigma^{-1}(n)}) \rangle_{M^n} \end{equation}
Therefore, 
\begin{equation} \sum_{\sigma} \operatorname{sgn}(\sigma) \langle \partial_{\theta_{\sigma(1)}} \ldots \partial_{\theta_{\sigma(n)}} F(\theta_1, \ldots, \theta_n) \rangle_{M^n} = \sum_{\sigma} \operatorname{sgn}(\sigma) \langle \partial_{\theta_1} \ldots \partial_{\theta_n} F(\theta_{\sigma^{-1} (1)}, \ldots, \theta_{\sigma^{-1}(n)}) \rangle_{M^n} \nonumber \end{equation}
\begin{equation}
= \sum_{\sigma} \operatorname{sgn}(\sigma)^2 \langle \partial_{\theta_1} \ldots \partial_{\theta_n} F(\theta_{1}, \ldots, \theta_{n}) \rangle_{M^n} 
= n! \langle \partial_{\theta_1} \ldots \partial_{\theta_n} F(\theta_1, \ldots, \theta_n) \rangle_{M^n}
\end{equation}
where we used the fact that $\sgn(\sigma) = \sgn(\sigma^{-1})$ for the second 
equality,  we replaced $\sigma^{-1}$ with $\sigma$
for the fourth term since summing over all $\sigma$ is the same as summing over all $\sigma^{-1}$, and used the fact that $F$ is fully antisymmetric for the 
fourth equality.   With this case established, we have completed the 
proof of Equation~\ref{Equivalence1} and thereby the proof of this 
theorem.
\end{proof}

\subsection{Emergence of Berezin multiple integrals from surface integrals as $D \rightarrow \infty$}
\label{highdimlimit} 

Now that we have shown the equivalence between the two types of multiple integrals, let us use Equation~\ref{DotDerivative} to evaluate their common value. Once again, our only concern is
with analytic function, or, in other words, multilinear functions 
of the noncommuting variables
$\theta_!, \ldots, \theta_n$.
Using the multilinearity and 
Equation~\ref{DotDerivative},
we see that it suffices to carry
out the calculation of 
$\int d \theta_1 * \cdots * d  \theta_n *(
\theta_1 \wedge \cdots \wedge 
\theta_n)$.  

We compute: 
\begin{equation}
\bigg \langle 
\partial_{\theta_1} 
\cdots \partial_{\theta_n}
\theta_1 \wedge \cdots 
\wedge \theta_n \bigg \rangle 
= \frac{1}{D^n} \sum_{\ell_1 = 1, \ldots, \ell_n = 1}^D e_{\ell_1} *
\cdots * e_{\ell_n} *
(e_{\ell_1} \wedge \cdots 
\wedge e_{\ell_n} )\nonumber
\end{equation}
\begin{equation} 
= \frac{1}{D^n} \sum_{\ell_1=1, 
\ldots, \ell_n = 1, \ell_j \neq \ell_k \text{ for $j \neq k$ } }^D 
e_{\ell_1} * \cdots 
* e_{\ell_n} * 
(e_{\ell_1} \wedge 
\cdots \wedge_{\ell_n}) 
\nonumber
\end{equation}
\begin{equation}
= \frac{1}{D^n} 
\sum_{\sigma \in S_n} 
\sum_{1 \leq \ell_1 < 
\cdots < \ell_n \leq D } 
\sgn(\sigma) e_{\ell_1} 
* \cdots *e_{\ell_n} 
* (\sgn(\sigma) 
e_{\ell_1} \wedge 
\cdots \wedge e_{\ell_n}) 
\nonumber \end{equation}
\begin{equation}
= \frac{n!}{D^n}
\sum_{1 \leq 1 \leq \ell_1 
< \cdots < \ell_n \leq D} 
e_{\ell_1} * \cdots 
* e_{\ell_n} * 
e_{\ell_1} * \cdots * 
e_{\ell_n} 
\nonumber
\end{equation}
\begin{equation}
= \frac{n!}{D^n} \binom{D}{n} (-1)^{n+1} 
= (-1)^{n+1} \frac{ \prod_{j=0}^{n-1} (D - j)}{D^n} 
\nonumber
\end{equation}

Using this calculation 
and specializing Equation~\ref{DotDerivative}
to the case of $F = 
\theta_1 \wedge 
\cdots \wedge \theta_n$, 
we conclude that for 
any choice of a region of
integration 
$M \subseteq V_D$ we have 
the following.

\begin{equation}
\int_{\partial M^n} d \theta_1 * \cdots d * \theta_n *
(\theta_1 \wedge \cdots \wedge 
\theta_n) = 
(\operatorname{vol}(M) D)^n
\bigg \langle 
\partial_{\theta_1} 
\cdots \partial_{\theta_n} 
\theta_1 \wedge \cdots \wedge
\theta_n \bigg \rangle_{M^n}
\label{Coefficient}
\end{equation}
\begin{equation}
= (\operatorname{vol}(M) D)^n
(-1)^{n+1} \frac{\prod_{j=0}^{n-1} (D-j)}{D^n} 
= (-1)^{n+1} \operatorname{vol}(M)^n
\prod_{j=0}^{n-1} (D - j)
\nonumber
\end{equation}

Thus, if we take $M \subseteq 
V_D$ to have $\operatorname{vol}
(M) = \frac{1}{D}$, and 
let $D$ tend to infinity, we
obtain the following.

\begin{equation}
\lim_{D \to \infty, 
\operatorname{vol}(M) = \frac{1}{D}}  \int_{\partial M^n} 
d \theta_1 * \cdots * d \theta_n * 
(\theta_1 \wedge \cdots \wedge 
\theta_n)  = (-1)^{n+1}
\end{equation}

With the sign conventions 
of~\cite{Drees}, this limit 
matches the formal computations
of iterated Berezin integrals.

\subsection{Minimalist ($D=n$) approach} \label{Misc1}

In Section~\ref{highdimlimit}, we realized 
iterated Berezin integrals as high dimensional
limits of geometric integrals.  
Intuitively, the purpose of moving to these
high dimensional spaces was to ensure that
$n$ randomly selected vectors be statistically
orthogonal.  If we were willing to 
simply enforce orthogonality by suitably
restricting the domain of integration, 
then we may also see the Berezin integrals
as geometric integrals.

For this to work, we still need $D \geq n$,
but the computation may be completed with $D = n$.  We constrain $\theta_k$ to the line segment 
$$M_k :=  \bigg\{ r e_k ~:~ \frac{-1}{2} 
\leq r \leq \frac{1}{2} \bigg\}$$

so that now the boundary of $M_k$ is the
set 

$$\partial M_k =  \bigg\{ \frac{-1}{2} e_k, \frac{1}{2} e_k \bigg\} \text{ .}$$

We compute immediately that 
$$
\int_{\partial M_k} d \theta_k = 
- e_k + e_k = 0
$$

and 

$$
\int_{\partial M_k} d\theta_k * \theta_k 
= (-e_k) * \frac{-e_k}{2} + e_k * \frac{e_k}{2} 
= 1
$$

Working with multiple integrals, 
we compute

\begin{equation} \int_{\partial_{M_1} \times \ldots \times \partial_{M_n} } (d \theta_1 \wedge \ldots \wedge d \theta_n) * (\theta_1 \wedge \ldots \wedge \theta_n)  \nonumber \end{equation}
\begin{equation} = \sum_{s_1=1}^2 \ldots \sum_{s_n=1}^2 (((-1)^{s_1} e_1) \wedge \ldots \wedge ((-1)^{s_n} e_n)) * \bigg( \bigg(\frac{(-1)^{s_1}}{2} e_1 \bigg) \wedge \ldots \wedge \bigg(\frac{(-1)^{s_n}}{2} e_n \bigg) \bigg)  \nonumber \end{equation}
\begin{equation} = \sum_{s_1=1}^2 \ldots \sum_{s_n=1}^2 \frac{1}{2^n} 
e_1 *\cdots *e_n  *e_1 \cdots *e_n = 2^n \frac{1}{2^n} (-1)^{n+1} = (-1)^{n+1} \end{equation}

More generally, for $f(\theta_1,
\ldots, \theta_n)$ an analytic 
function of the noncommuting variables
$\theta_1, \ldots, \theta_n$, 
using multilinearity and the
fact that our constraints enforce
orthogonality so that 
$d \theta_1 \cdots d \theta_n = 
d \theta_1 \wedge \cdots \wedge 
d \theta_n$, we see that 

\begin{equation}
\int_{\partial M_1 \times \cdots 
\times \partial M_n} d \theta_1 *
\cdots * d \theta_1 *f(\theta_1, 
\ldots, \theta_n) = 
\int_{\partial M_1 \times \cdots 
\times \partial M_n} (d \theta_1 \wedge
\cdots \wedge d \theta_1) * f(\theta_1, 
\ldots, \theta_n)  \nonumber \end{equation}
\begin{equation} = 
\int^{\text{Ber}} d \theta_1 \cdots d \theta_n
f(\theta_1, \ldots, \theta_n)
\end{equation}

\subsection{Enforcing orthogonality with intermediate values of $D>n$}  \label{Misc2}

In the previous section we saw that, if we set a constraint that the vectors are orthogonal, we no longer have to set $D$ to infinity and, in fact, $D=n$ can suffice. However, if there are other reasons for which we want $D$ to be larger than $n$, this is also allowed. For example, if we wish for $\theta$ to have continuous spectrum, we would set $D=2n$. That would result in any given variable $\theta_k$ being confinded to a curve $C_k$, which would be ``$1$-dimensional hypersurface". Its ``area element" would be the length, yet it would be perpendicular to the direction of a curve. This, of course, is rather odd. A situation that would have looked a lot more usual would of been if instead of having one-dimensional hypersurface embedded in 2-dimensional space we had 2-dimensional hypersurface embedded into 3 dimensional space. For that, we would have to set $D=3n$ and confine $\theta_k$ to a closed surface in $(3k-2, 3k-1, 3k)$-hyperplane that encloses the volume $1/3$. And, finally, we for the sake of generality, we may leave $D \geq n$, or $D \geq 2n$ or $D \geq 3n$ undefined via having greater or equal signs rather than equal signs. 

\subsection{Directed volume measure} \label{DirectedVolume}

One problem that the above approaches have is the one of rotational invariance. One can show that if $\theta_1$ and $\theta_2$ are both confined to $D-1$ dimensional hypersurface then $(\theta_1 + \theta_2)/ \sqrt{2}$ would span a $D$-dimensional region, which would put it into a different footing from $\theta_1$ and $\theta_2$. If we take minimalist approach, then $\theta_1$ and $\theta_2$ would each take two values, while $(\theta_1 + \theta_2)/ \sqrt{2}$ would take four values, which again puts it in a different footing. We propose to resolve this issue by saying that each $\theta_1$ and $\theta_2$ spans $D$ dimensional space rather than $D-1$, which would put it on the same footing as $\theta_1+ \theta_2$. In case of area element, a crucial point was that it is vector-valued. Thus, we need the volume element to be vector-valued as well. In particular, we define a function $\mu \colon V_D \mapsto \mathbb{R}$ and then define volume element to be 
\begin{equation} d_{\mu} \theta = \frac{\theta}{\vert \theta \vert} \mu (\theta) dV \label{VolumeMeasure} \end{equation} 
From this we obtain
\begin{equation} \int d_{\mu} \theta = \int \frac{\theta}{\vert \theta \vert} \mu (\theta) dV \end{equation}
\begin{equation} \int \theta * d_{\mu} \theta = \int \frac{\theta * \theta}{\vert \theta \vert} \mu (\theta) dV = \int \frac{\vert \theta \vert^2}{\vert \theta \vert} \mu (\theta) dV = \int \vert \theta \vert \mu (\theta) dV \end{equation}
If, in particular, we make $\mu$ symmetric, 
\begin{equation} \mu (\theta) = \mu (\vert \theta \vert) \end{equation}
then from rotational symmetry 
\begin{equation} \int d_{\mu} \theta = \int \frac{\theta}{\vert \theta \vert} \mu (\vert \theta \vert) dV =0 \end{equation}
and, if we denote the area of the unit sphere in $D$ dimensions by $a_D$, we get 
\begin{equation} \int \theta * d_{\mu} \theta = \int \vert \theta \vert \mu (\vert \theta \vert) dV = a_D \int_0^{\infty} r^D \mu (r) dr \label{Norm} \end{equation}
The above can be made $1$ by normalizing $\mu$ in the following way: 
\begin{equation} \mu (r) = \frac{\mu_0 (r)}{a_D \int_0^{\infty} r^D \mu_0 (r) dr} \label{NormalizationVolume} \end{equation}
As it turns out, if $\rho$ is symmetric, there is a relation between volume integral and surface integral. First of all, we can split volume integral into the surface integrals by confining each $\theta_k$ to its own sphere of radius $r_k$, evaluating the surface integral, and then taking the integral of our answer over all possible $r_1, \cdots, r_n$: 
\begin{equation} \int (d_{\rho} \theta_1 \cdots d_{\rho} \theta_n) * f(\theta_1, \cdots, \theta_n) = \label{Slicing} \end{equation}
\begin{equation} =  \int \bigg( dr_1 \cdots dr_n \rho (r_1) \cdots \rho (r_n) \int_{S(r_1) \times \cdots \times S(r_n)} (d \theta_1 \cdots d \theta_n) * f(\theta_1, \cdots, \theta_n) \bigg) \nonumber  \end{equation}
where we skipped product sings in $d \theta_1 \cdots d \theta_n$ to designate that this reasoning will go through regardless of which product sign is inserted. We then note that 
\begin{equation} \int_{S(r_1) \times \cdots \times S(r_n)} (d \theta_1 \cdots d \theta_n) * f(\theta_1, \cdots, \theta_n)  \nonumber \end{equation}
\begin{equation} =  r^{D-1}_1 \cdots r^{D-1}_n \int_{S(1) \times \cdots \times S(1)} (d \theta_1 \cdots d \theta_n) * f(r_1\theta_1, \cdots, r_n\theta_n) \end{equation}
\emph{If we assume $f$ is analytic} with respect to $\wedge$-product, then the only term of $f$ that would not drop out of the integral would be the one proportional to $\theta_1 \wedge \cdots \wedge \theta_n$ which would provide an extra coefficient of $r_1 \cdots r_n$, leading to
\begin{equation} (r_1 \cdots r_n )(r_1^{D-1} \cdots r_n^{D-1}) = r_1^D \cdots r_n^D \end{equation}
Thus, 
\begin{equation} \int_{S(r_1) \times \cdots \times S(r_n)} (d \theta_1 \cdots d \theta_n) * f (\theta_1, \cdots \theta_n) = \end{equation}
\begin{equation} = r_1^D \cdots r_n^D \int_{S(1) \times \cdots \times S(1)} (d \theta_1 \cdots d \theta_n) * (\theta_1 \wedge \cdots \wedge \theta_n) \nonumber \end{equation}
By substituting this into Eq \ref{Slicing}, we obtain 
\begin{equation} \int (d_{\rho} \theta_1 \cdots d_{\rho} \theta_n) * f(\theta_1, \cdots, \theta_n) = \end{equation}
\begin{equation} \int \bigg( dr_1 \cdots dr_n \rho (r_1) \cdots \rho (r_n) r_1^D \cdots r_n^D \int_{S(1) \times \cdots \times S(1)} (d \theta_1 \cdots d \theta_n) * f(\theta_1 \wedge \cdots \wedge \theta_n) \bigg)  \nonumber  \end{equation}
\begin{equation} = \bigg( \int dr_1 \cdots dr_n \rho (r_1) \cdots \rho (r_n) r_1^D \cdots r_n^D \bigg) \bigg(\int_{S(1) \wedge \cdots \wedge S(1)} (d \theta_1 \wedge \cdots \wedge \theta_n) f(\theta_1, \cdots, \theta_n) \bigg)  \nonumber \end{equation}
\begin{equation} = \bigg(\int dr \rho (r) r^D \bigg)^n \bigg(\int_{S(1) \wedge \cdots \wedge S(1)} (d \theta_1 \wedge \cdots \wedge \theta_n) \bigg)  \nonumber \end{equation}
This result implies that the equations for surface integrals can be carried over for volume integrals. For example, from Eq \ref{Coefficient} for a surface integral, we obtain the volume integral expression
\begin{equation} \int d_{\mu} \theta_1 * \ldots *d_{\mu} \theta_n * (\theta_1 \wedge \ldots \wedge \theta_n)  \nonumber \end{equation}
\begin{equation} = \int (d_{\mu} \theta_1 \wedge \ldots \wedge d_{\mu} \theta_n)*(\theta_1 \wedge \ldots \wedge \theta_n) = (-1)^{n+1} \frac{\prod_{j=0}^{n-1} (D-j)}{D^n} \int d_{\mu} \theta * \theta \end{equation}
which also becomes $1$, provided we normalize $\mu$ via Eq \ref{NormalizationVolume} and take limit of $D \rightarrow \infty$. 

Let us see which choices of $\mu$ would result in the rotational invariance. Consider $n$ variables $\theta_1, \ldots, \theta_n$, each living in the same space $V_D$ (thus, there are total of $nD$ real coordinates). Consider the change of variables 
\begin{equation} \theta^{\prime}_i = \sum_{j=1}^n A_{ij} \theta_j \end{equation}
The criteria for rotational invariance can be stated as follows: 
\begin{equation} {\rm Rotational \; Invariance} \Longleftrightarrow d_{\mu} \theta^{\prime}_1 \wedge \ldots \wedge d_{\mu} \theta^{\prime}_n = (\det A) d_{\mu} \theta_1 \wedge \ldots \wedge d_{\mu} \theta_n \end{equation}
Now, the left and right hand side of the above evaluate to
\begin{equation} LHS= d_{\mu} \theta^{\prime}_1 \wedge \ldots \wedge d_{\mu} \theta^{\prime}_n = \prod_{i=1}^n \frac{\theta^{\prime}_i}{\vert \theta^{\prime}_i \vert} \mu (\vert \theta^{\prime}_i \vert) = (\det A) \prod_{i=1}^n \frac{\theta_i}{\vert \theta_i^{\prime} \vert} \mu (\vert \theta_i^{\prime} \vert) \end{equation}
\begin{equation} RHS = (\det A) d_{\mu} \theta_1 \wedge \ldots \wedge d_{\mu} \theta_n = (\det A) \prod_{i=1}^n \frac{\theta_i}{\vert \theta_i \vert} \mu (\vert \theta_i \vert) \end{equation} 
From this, we have 
\begin{equation} {\rm Rotational \; Invariance} \Longleftrightarrow \prod_{i=1}^n \frac{\mu (\vert \theta_i^{\prime} \vert)}{\vert \theta_i^{\prime} \vert} = \prod_{i=1}^n \frac{\mu (\vert \theta_i \vert)}{\vert \theta_i \vert} \end{equation} 
That is being satisfied by 
\begin{equation} \frac{\mu (\vert \theta \vert)}{\vert \theta \vert} = k_D e^{- \alpha \vert \theta \vert^2/2} \end{equation}
which corresponds to 
\begin{equation} d_{\mu} \theta = k_D \theta e^{- \alpha \vert \theta \vert^2/2} dV \label{DirectedMeasure} \end{equation}
where the product sign after $\theta$ was skipped in light of the fact that $e^{- \alpha \vert \theta \vert^2/2}$ is a real number. To find coefficient $k_D$, we compute the integral of $\theta * d_{\mu} \theta$ and then find the value of $k_D$ that would set it to $1$: 
\begin{equation} \int \theta * d_{\mu} \theta = k_D \int \vert \theta \vert^2 e^{- \alpha \vert \theta \vert^2/2} dV= k_D \sum_{i=1}^D \int x_i^2 e^{- \frac{\alpha}{2} (x_1^2+ \cdots + x_D^2)} dx_1 \cdots dx_D = \end{equation} 
\begin{equation} = k_D \sum_{i=1}^D \bigg(\int x_i^2 e^{- \alpha x_i^2/2} dx_i \bigg) \bigg(\prod_{j \neq i}  \int e^{- \alpha x_j^2/2} dx_j \bigg)  \nonumber \end{equation}
\begin{equation} = \frac{k_D}{\alpha^{1+ \frac{D}{2}}} \sum_{i=1}^D \bigg(\int y_i^2 e^{-y_i^2/2}dy_i \bigg) \bigg(\prod_{j \neq i} \int e^{-y_j^2/2} dy_j \bigg)  \nonumber \end{equation} 
\begin{equation} = \frac{k_D}{\alpha^{1+ \frac{D}{2}}} \sum_{i=1}^D (\sqrt{2 \pi}) (\sqrt{2 \pi})^{D-1} =  \frac{k_D}{\alpha^{1+ \frac{D}{2}}} \sum_{i=1}^D (2 \pi)^{D/2} = \frac{k_DD (2 \pi)^{D/2}}{\alpha^{1+ \frac{D}{2}}} \nonumber \end{equation}
If we set the above integral to $1$, we get
\begin{equation} k_D = \frac{\alpha^{1+ \frac{D}{2}}}{D (2 \pi)^{D/2}} \label{kReal} \end{equation}
It should be noted, however, that the above was done for the measure $d_{\mu} \theta_1 \wedge \ldots \wedge d_{\mu} \theta_n$, as opposed to $d_{\mu} \theta_1 \ldots d_{\mu} \theta_n$. In the latter case, we simply point to the fact proven earlier that integration with respect to either measure gives the same result due to the fact that the difference between two measures integrates to zero. 

\subsection{Complex Grassmann numbers and spinors} \label{ComplexGrassmannNumbers}

One issue about complex Grassmann numbers that needs to be addressed is that, if $d \theta$ were parallel to $\theta$, then under the transformation $\theta \mapsto i \theta$ we would have $d \theta \mapsto i d \theta$ and, therefore, $d \theta * \theta \mapsto - d \theta * \theta$. If we were to integrate $d \theta * \theta$ over the whole complexified space then, from the above, we would expect the integral to be equal to minus itself and, therefore, zero. In order to avoid this, we say that $d \theta$ is parallel to $\theta^*$ rather than $\theta$ (where $*$ denotes complex conjugation), which we will denote by writing $d_{\mu^*} \theta$ as opposed to $d_{\mu} \theta$. Thus, 
\begin{equation} d_{\mu^*} \theta = k_{2D} \theta^* e^{- \alpha \vert \theta \vert^2/ 2} dV \end{equation}
which would allow for the integral over $d_{\mu^*} \theta * \theta$ to produce a non-zero result, and $k_{2D}$ is given by Eq \ref{kReal}, where $D$ is being replaced by $2D$ to signify that $D$ complex dimensions is the same as $2D$ real dimensions; thus, 
\begin{equation} k_{2D} = \frac{\alpha^{1+D}}{2D (2 \pi)^D} \end{equation} 
Going back to our discussion about conjugation, just as $d_{\mu} \theta$ is parallel to $\theta^*$ rather than $\theta$, similarly $d_{\mu^*} \theta^*$ will be parallel to $\theta$ rather than $\theta^*$, 
\begin{equation} d_{\mu^*} \theta^* = k_D \theta e^{- \alpha \vert \theta \vert^2/ 2} dV \end{equation}
which is the reason why $d_{\mu^*} \theta^* * \theta$ would integrate to zero. Sometimes, for the sake of brevity, we will skip $\mu$ or $\mu^*$ below $d$. In this case, it should be understood that 
\begin{equation} d (\cdots) = d_{\mu^*} (\cdots) \end{equation}

In order to explain the symmetry properties of spinors, we should define spinors as a linear combinations of \emph{tensor products} between unit vectors $\vert v \rangle \in \mathbb{C}^4$ and Grassmann variables $\psi \in {\mathcal V}_D$:
\begin{equation} \psi = \sum_{k=1}^4 \psi_k \otimes \vert v_k \rangle \end{equation}
which is to be contrasted with a commuting spinor, given by 
\begin{equation} \vert v \rangle = \sum_{k=1}^4 c_k \vert v_k \rangle \; , \; c \in \mathbb{C} \end{equation}
We can define the Lorentz transformation to be 
\begin{equation} \psi^{\prime}_k = \sum_{i=1}^4  \psi_i (U^{\dagger})_{ik} \end{equation}
\begin{equation} \vert v_k \rangle = \sum_{j=1}^4  U_{kj} \vert v_j \rangle \end{equation}
and, therefore, 
\begin{equation} \sum_{k=1}^4 \psi^{\prime}_k \vert v^{\prime}_k \rangle = \sum_{i=1}^4 \sum_{j=1}^4 \sum_{k=1}^4 \psi_i (U^{\dagger})_{ik} U_{kj} \vert v_j \rangle = \sum_{i=1}^4 \psi_i \vert v_i \rangle \end{equation}
The measure on $\psi = (\psi_1, \psi_2, \psi_3, \psi_4)$ is
\begin{equation} d \psi =  k^4 \psi_1^* \psi_2^* \psi_3^* \psi_4^* e^{- \alpha \vert \psi \vert^2/2} dV \end{equation}
where
\begin{equation} \vert \psi \vert^2 = \vert \psi_1 \vert^2 + \vert \psi_2 \vert^2 + \vert \psi_3 \vert^2 + \vert \psi_4 \vert^2 \end{equation} 
In light of the fact that $U$ is unitary, one can show that 
\begin{equation} \psi_1^{\prime *}  \psi_2^{\prime *}  \psi_3^{\prime *}  \psi_4^{\prime *} = \psi_1^*  \psi_2^*  \psi_3^*  \psi_4^* \end{equation}
\begin{equation} \vert \psi_1^{\prime} \vert^2 + \vert \psi_2^{\prime} \vert^2 + \vert \psi_3^{\prime} \vert^2 + \vert \psi_4^{\prime} \vert^2 =  \vert \psi_1 \vert^2 + \vert \psi_2 \vert^2 + \vert \psi_3 \vert^2 + \vert \psi_4 \vert^2 \end{equation}
which implies that the measure of integration on $\psi$ is the same as the measure of integration in $\psi^{\prime}$. Therefore, as long as the domain of integration itself is symmetric under the spin transformation, so are the integrals. We will end this section by mentioning two ways in which it could have been written differently: 

1) Instead of using a volume integral we could have used a surface integral. Then, as discussed earlier, we would have to select a preferred reference frame. That is because under a coordinate transformation the surface integral becomes the volume integral, thus breaking the symmetry; but if we start from the volume integral as given above, it will retain its form, thus symmetry would be respected. The violation of spin symmetry has no consequence on actual physical results, as long as the Lagrangian respects that symmetry. Therefore, it is really an aesthetic question. The reason to use surface integrals is that the concept of directed area element is standard while the concept of directed volume is not; the reason to use volume integral is that the spin symmetry is respected this way. 
 
2) In \cite{Book} it was suggested to interpret spinors themselves in terms of geometric algebra. However, that approach is different from ours since they identify anticommuting units with $\sigma_k$, and there are three of them, in contrast to us having $D \rightarrow \infty$ components.  Apart from that, they define spinors to be even rank objects, while we have to have odd rank in order to explain anticommutativity (more specifically, we define them to be rank 1, although in principle we could have made them more general odd rank, which we chose not to do). Both of those observations point to the fact that they are focused on re-interpreting $\vert v_k \rangle$ as opposed to $e_k$. If we substitute their definition of $\vert v_k \rangle$ into $e_k \otimes \vert v_k \rangle$, we would get a tensor product between our structure and theirs. Since this is isomorphic to what we already wrote, it would neither help nor hinder us, so we will leave this out. However, a direction of research that might be worthwhile is to modify what they have in such a way that tensor product would no longer be necessary and, instead, we would obtain \emph{both} $\sigma$-matrices \emph{and} $e_k$-s from a single underlying framework. But, as of now, we are not sure how to do that, so we leave it for a future research. 

\section{Glimpse of supersymmetry} \label{SUSY} 

\subsection{Orbit of $\mathbb{R}^4$ under supersymmetric transformations} \label{Orbit}

It is not our intention to provide a full supersymmetric theory (which deserves a separate paper); our only goal here is to show that in principle its key aspects translate into our framework. On the positive side, since we have reinterpreted Grassmann numbers as literal elements of a continuous space, superspace should admit an interpretation as a continuous space as well in the literal sense of the word. However, there are some basic questions that we need to resolve. For example, superspace transformation would add a rank 2 component to space time points $x \in \mathbb{R}^4$. Thus, the spacetime would no longer be $\mathbb{R}^4$ but, instead, it would have to be extended to include rank-$2$ objects. What would this new space be? Also, when we are taking integrals over a superpotential to get a usual action, what subspace of that superspace are we integrating over? In the section that follows, we will mainly focus on those two questions, and we will restrict ourselves to only a couple of the simplest examples. We hope to convince the reader that one could proceed in similar fashion to ``translate" other aspects of supersymmetry into our framework as well, but explicitly doing so is beyond the scope of this paper. 

The supersymmetry transformation is generated through infinitesimal transformation, of the form
\begin{equation} \theta^{\alpha \prime} = \theta^{\alpha} -i \epsilon^{\alpha} \; , \; \overline{\theta}^{\dot{\alpha} \prime} = \overline{\theta}^{\dot{\alpha}} + i\overline{\epsilon}^{\dot{\alpha}} \; , \; x^{\mu \prime} = x^{\mu} + \epsilon^{\alpha} \sigma^{\mu}_{\alpha \dot{\beta}} \wedge \overline{\theta}^{\dot{\beta}}  - \overline{\epsilon}^{\dot{\alpha}} \wedge  \theta^{\beta} \sigma^{\mu}_{\beta \dot{\alpha}} \label{SupersymmetryTransformation} \end{equation}
where $\epsilon^{\alpha}$ and $\epsilon^{\dot{\alpha}}$ are infinitesimal rank-1 objects with spinor indexes, $\sigma$-matrices are given by 
\[ \sigma^0 = \left( \begin{array}{cc}
1 & 0 \\
0 & 1 \end{array} \right) \; \; \;  \sigma^1 = \left( \begin{array}{cc} 0 & 1  \\
1 & 0 \end{array}  \right) \; \; \; \sigma^2 = \left( \begin{array}{cc}
0 & -i  \\
i & 0 \end{array} \right) 
\; \; \; \sigma^3 = \left( \begin{array}{cc}
1 & 0 \\
0 & -1 \end{array} \right) \] 
and \emph{Einsten summation convention} is used; that is, repeated indexes are summed over. The above transformations are generated by objects of the type $\epsilon^{\alpha} \wedge Q_{\alpha}$ and $\epsilon^{\dot{\alpha}} \wedge Q_{\dot{\alpha}}$, where $\epsilon^{\alpha}$ and $\epsilon^{\dot{\alpha}}$ are infinitesimal rank-1 objects while $Q_{\alpha}$ and $Q_{\dot{\alpha}}$ are \emph{supercharges}, defined as 
\begin{equation} Q_{\alpha} = -i (\partial_{\alpha} +i \sigma^{\mu}_{\alpha \dot{\beta}} \overline{\theta}^{\dot{\beta}} \partial_{\mu}) \label{Q} \end{equation} 
\begin{equation} \overline{Q}_{\dot{\alpha}} = i (\overline{\partial}_{\dot{\alpha}} + i \theta^{\beta} \overline{\sigma}^{\mu}_{\beta \dot{\alpha}} \partial_{\mu}) \label{QOverline}\end{equation}
In the above expressions, as well as in the future, it is understood that the indexes from the beginning of the alphabet, such as $\alpha$, $\beta$, $\gamma$ or $\delta$ refer to $\theta$, while the indexes from the middle of the alphabet, such as $\mu$, $\nu$, $\rho$ and $\sigma$ refer to $x$. Thus, 
\begin{equation} \partial_{\alpha} = \frac{\partial}{\partial \theta^{\alpha}} \end{equation}
\begin{equation} \overline{\partial}_{\dot{\alpha}} = \frac{\partial}{\partial \overline{\theta}^{\dot{\alpha}}} \end{equation}
\begin{equation} \partial_{\mu} = \frac{\partial}{\partial x^{\mu}} \end{equation}
In light of the fact that we have more than one product, we define commutators and anticommutators for each product. First, we define the products of operators. For an operator $A$, define operators $\theta \wedge A$, $\theta *A$, $\partial_{\theta} \wedge A$ and $\partial_{\theta} * A$ as 
\begin{equation} (\theta \wedge A) {\mathcal F} = \theta \wedge (A {\mathcal F}) \; , \; (\theta *A) {\mathcal F} = \theta *(A {\mathcal F}) \; , \; (\partial_{\theta} \wedge A) {\mathcal F} = (\partial_{\theta}* A) {\mathcal F} = \partial_{\theta} (A {\mathcal F}) \end{equation} 
where $\mathcal F$ is some test superfield those operators are acting on. We then define commutation and anticommutation relations as 
\begin{equation} [A,B]_*= A*B-B*A \; , \; [A,B]_{\wedge} = A \wedge B - B \wedge A \end{equation}
\begin{equation} \{A, B \}_* = A*B+B*A \; , \; \{A,B\}_{\wedge} = A \wedge B + B \wedge A \end{equation} 
With this notation, the supercharges satisfy anticommutation relations 
\begin{equation} \{Q_{\alpha}, Q_{\beta} \}_{\wedge} = \{\overline{Q}_{\dot{\alpha}}, \overline{Q}_{\dot{\beta}} \}_{\wedge} = 0 \end{equation} 
\begin{equation} \{Q_{\alpha}, \overline{Q}_{\dot{\beta}} \}_{\wedge} = 2i \sigma^{\mu}_{\alpha \dot{\beta}} \partial_{\mu} \end{equation}

Inspection of the equations for $Q$ and $\overline{Q}$ tells us four things. On the one hand, commuting components can not be confined to $\mathbb{R}^4$ since $Q$ and $\overline{Q}$ produce elements of  $\Big( \bigwedge^2 (V_D) \Big)_{\mathbb{C}} \otimes_{\mathbb{C}} (T \mathbb{R}^4)_{\mathbb{C}}$ that would also be commuting, although not real. On the other hand, commuting components \emph{do} stay inside of $\mathbb{R}^4 +  \Big( \bigwedge^2 (V_D) \Big)_{\mathbb{C}} \otimes_{\mathbb{C}} (T \mathbb{R}^4)_{\mathbb{C}}$. Thirdly, the transformations do not affect the rank-0 component, since those transformations acting on rank 0 produce rank 1, while their action on rank 1 produces rank 2. And, finally, the transformations are independent of the rank 0 component since they depend on rank 1 elements $\epsilon^{\alpha}$ and $\overline{\epsilon}^{\dot{\alpha}}$.  So our next question is: what subset of $\mathbb{R}^4 + \Big( \bigwedge^2 (V_D) \Big)_{\mathbb{C}} \otimes_{\mathbb{C}} (T \mathbb{R}^4)_{\mathbb{C}}$ would be generated by $Q$ and $\overline{Q}$? There is a related question: when we will be integrating a Lagrangian to get an action, what set would we be taking integral over and what would be its measure? 
 
\begin{thm} The orbit of any given $g \in \mathbb{R}^4 + ((V_D)_{\mathbb{C}})^4 + \Big( \bigwedge^2 (V_D) \Big)_{\mathbb{C}} \otimes_{\mathbb{C}} (T \mathbb{R}^4)_{\mathbb{C}}$ is all of $ g + ((V_D)_{\mathbb{C}})^4 + \Big( \bigwedge^2 (V_D) \Big)_{\mathbb{C}} \otimes_{\mathbb{C}} (T \mathbb{R}^4)_{\mathbb{C}}$.
\end{thm}
\begin{proof} We obtain from Eq \ref{SupersymmetryTransformation}, \ref{Q} and \ref{QOverline} that $(g+ (V_D)_{\mathbb{C}})^4$ is contained in the orbit of $g$ under the action of operators of the form $\xi^{\alpha} \wedge Q_{\alpha}$ and $\overline{\eta}^{\dot{\alpha}} \wedge \overline{Q}_{\dot{\alpha}}$. We shall show now that commutators of such operators generate the space $(\bigwedge^2 (V_D))_{\mathbb{C}} \otimes_{\mathbb{C}} (T \mathbb{R})^4)_{\mathbb{C}}$.  First note that the multiplication by anticommuting elements turns commutator into an anticommutator
\begin{equation} [\xi^{\alpha} \wedge Q_{\alpha}, \overline{\eta}^{\dot{\beta}} \wedge \overline{Q}_{\dot{\beta}}]_{\wedge}  = \xi^{\alpha} \wedge \overline{\eta}^{\dot{\beta}} \{ Q_{\alpha}, \overline{Q}_{\dot{\beta}} \}_{\wedge} \end{equation}
Therefore, 
\begin{equation} [\xi^{\alpha} \wedge Q_{\alpha}, \overline{\eta}^{\dot{\beta}} \wedge \overline{Q}_{\dot{\beta}}]_{\wedge} = 2i \sigma^{\mu}_{\alpha \dot{\beta}} \xi^{\alpha}  \wedge \overline{\eta}^{\dot{\beta}} \partial_{\mu} \end{equation}
By trying out different ways of substituting $+e_i$, $-e_i$, $+ie_i$ and $-ie_i$ into $\xi$ and $\overline{\eta}$ we obtain 
\begin{equation} [e_i \wedge (Q_1 + Q_2), e_j \wedge (\overline{Q}_{\dot{1}} + \overline{Q}_{\dot{2}})]_{\wedge} = 2i e_i \wedge e_j  (\sigma^{\mu}_{1 \dot{1}} + \sigma^{\mu}_{1 \dot{2}} + \sigma^{\mu}_{2 \dot{1}} + \sigma^{\mu}_{2 \dot{2}}) \partial_{\mu} = 4i e_i \wedge e_j (\partial_0 + \partial_1) \end{equation} 
\begin{equation} [e_i \wedge (Q_1 - Q_2), e_j \wedge (\overline{Q}_{\dot{1}} - \overline{Q}_{\dot{2}})]_{\wedge} = 2i e_i \wedge e_j  (\sigma^{\mu}_{1 \dot{1}} - \sigma^{\mu}_{1 \dot{2}} - \sigma^{\mu}_{2 \dot{1}} + \sigma^{\mu}_{2 \dot{2}}) \partial_{\mu} = 4i e_i \wedge e_j (\partial_0 - \partial_1) \end{equation}

\begin{equation} [e_i \wedge (Q_1 -i Q_2), e_j \wedge (\overline{Q}_{\dot{1}} +i \overline{Q}_{\dot{2}})]_{\wedge} = 2i e_i \wedge e_j  (\sigma^{\mu}_{1 \dot{1}} +i \sigma^{\mu}_{1 \dot{2}} -i \sigma^{\mu}_{2 \dot{1}} + \sigma^{\mu}_{2 \dot{2}}) \partial_{\mu} = 4i e_i \wedge e_j (\partial_0 + \partial_2) \end{equation}

\begin{equation} [e_i \wedge (Q_1 +i Q_2), e_j \wedge (\overline{Q}_{\dot{1}} -i \overline{Q}_{\dot{2}})]_{\wedge} = 2i e_i \wedge e_j  (\sigma^{\mu}_{1 \dot{1}} -i \sigma^{\mu}_{1 \dot{2}} +i \sigma^{\mu}_{2 \dot{1}} + \sigma^{\mu}_{2 \dot{2}}) \partial_{\mu} = 4i e_i \wedge e_j (\partial_0 - \partial_2) \end{equation} 
\begin{equation} [e_i \wedge Q_1, e_j \wedge \overline{Q}_{\dot{1}}]_{\wedge} = 2i e_i \wedge e_j \sigma^{\mu}_{1 \dot{1}} \partial_{\mu} = 2i e_i \wedge e_j (\partial_0 + \partial_3) \end{equation}
\begin{equation} [e_i \wedge Q_2, e_j \wedge \overline{Q}_{\dot{2}}]_{\wedge} = 2i e_i \wedge e_j \sigma^{\mu}_{2 \dot{2}} \partial_{\mu} = 2i e_i \wedge e_j (\partial_0 - \partial_3) \end{equation}
The linear combinations of the above equations give us 
\begin{equation}  e_i  \wedge e_j \partial_0 = - \frac{1}{4} [(ie_i) \wedge Q_1, e_j \wedge \overline{Q}_{\dot{1}}]_{\wedge} - \frac{1}{4} [ie_i \wedge Q_2, e_j \wedge \overline{Q}_{\dot{2}}]_{\wedge} \end{equation}
\begin{equation}  e_i \wedge e_j \partial_1 =  - \frac{1}{4} [(ie_i) \wedge Q_1, e_j \wedge \overline{Q}_{\dot{2}}]_{\wedge} - \frac{i}{4} [ie_i \wedge Q_2, e_j \wedge \overline{Q}_{\dot{1}}]_{\wedge} \end{equation}
\begin{equation}  e_i \wedge e_j \partial_2 = \frac{1}{4} [e_i \wedge Q_1, e_j \wedge \overline{Q}_{\dot{2}}]_{\wedge} - \frac{1}{4} [e_i \wedge Q_2, e_j \wedge \overline{Q}_{\dot{1}}]_{\wedge} \end{equation} 
\begin{equation}  e_i \wedge e_j \partial_3 = -  \frac{1}{4} [(ie_i) \wedge Q_1, e_j \wedge \overline{Q}_{\dot{1}}]_{\wedge} + \frac{1}{4} [(ie_i) \wedge Q_2, e_j \wedge \overline{Q}_{\dot{2}}]_{\wedge} \end{equation}
\begin{equation} i e_i  \wedge e_j \partial_0 =  \frac{1}{4} [e_i \wedge Q_1, e_j \wedge \overline{Q}_{\dot{1}}]_{\wedge} + \frac{1}{4} [e_i \wedge Q_2, e_j \wedge \overline{Q}_{\dot{2}}]_{\wedge} \end{equation}
\begin{equation} i e_i \wedge e_j \partial_1 =  \frac{1}{4} [e_i \wedge Q_1, e_j \wedge \overline{Q}_{\dot{2}}]_{\wedge} + \frac{1}{4} [e_i \wedge Q_2, e_j \wedge \overline{Q}_{\dot{1}}]_{\wedge} \end{equation}
\begin{equation}  i e_i \wedge e_j \partial_2 = \frac{1}{4} [(ie_i) \wedge Q_1, e_j \wedge \overline{Q}_{\dot{2}}]_{\wedge} - \frac{1}{4} [(ie_i) \wedge Q_2, e_j \wedge \overline{Q}_{\dot{1}}]_{\wedge} \end{equation} 
\begin{equation} i e_i \wedge e_j \partial_3 =  \frac{1}{4} [e_i \wedge Q_1, e_j \wedge \overline{Q}_{\dot{1}}]_{\wedge} - \frac{1}{4} [e_i \wedge Q_2, e_j \wedge \overline{Q}_{\dot{2}}]_{\wedge} \end{equation}
The reason we have $8$ equations is that for each $e_{\alpha}$ and $e_{\dot{\alpha}}$ we also considered $i e_{\alpha}$ and $ie_{\dot{\alpha}}$ since $\alpha$ is a spinor index and spinors are complex valued. This generates the whole complexified rank-2 component. 

Now, if we denote by $z^{\alpha i}$ and $\overline{z}^{\dot{\alpha} i}$ the real components of $\theta$ and $\overline{\theta}$, 
\begin{equation} \theta^{\alpha} = z^{\alpha i} e_i \end{equation}
\begin{equation} \overline{\theta}^{\dot{\alpha}} = \overline{z}^{\dot{\alpha} i} e_i \end{equation}
we can then do the following calculation 
\begin{equation} \label{DisplacementProof} e_i \partial_{\alpha} = e_i \wedge (iQ_{\alpha} - i \sigma^{\mu}_{\alpha \dot{\beta}} \overline{\theta}^{\dot{\beta}} \partial_{\mu} ) = i e_i \wedge Q_{\alpha} - i \sigma^{\mu}_{\alpha \dot{\beta}} \overline{z}^{\dot{\beta} j} e_i \wedge e_j \partial_{\mu} =  \end{equation}  
\begin{equation} =  i e_i \wedge Q_{\alpha} - i \sigma^0_{\alpha \dot{\beta}} \overline{z}^{\dot{\beta} j} \bigg( - \frac{i}{4} [e_i \wedge Q_1, e_j \wedge \overline{Q}_{\dot{1}}]_{\wedge} - \frac{i}{4} [e_i \wedge Q_2, e_j \wedge \overline{Q}_{\dot{2}}]_{\wedge} \bigg)  \nonumber \end{equation}
\begin{equation} - i \sigma^1_{\alpha \dot{\beta}} \overline{z}^{\dot{\beta} j} \bigg( - \frac{i}{4} [e_i \wedge Q_1, e_j \wedge \overline{Q}_{\dot{2}}]_{\wedge} - \frac{i}{4} [e_i \wedge Q_2, e_j \wedge \overline{Q}_{\dot{1}}]_{\wedge} \bigg)  \nonumber \end{equation}
\begin{equation} - i \sigma^2_{\alpha \dot{\beta}} \overline{z}^{\dot{\beta} j} \bigg(\frac{1}{4} [e_i \wedge Q_1, e_j \wedge \overline{Q}_{\dot{2}}]_{\wedge} - \frac{1}{4} [e_i \wedge Q_2, e_j \wedge \overline{Q}_{\dot{1}}]_{\wedge} \bigg)  \nonumber \end{equation}
\begin{equation} - i \sigma^3_{\alpha \dot{\beta}} \overline{z}^{\dot{\beta} j} \bigg( - \frac{i}{4} [e_i \wedge Q_1, e_j \wedge \overline{Q}_{\dot{1}}]_{\wedge} + \frac{i}{4} [e_i \wedge Q_2, e_j \wedge \overline{Q}_{\dot{2}}] _{\wedge} \bigg) \nonumber \end{equation}

\begin{equation} e_i \overline{\partial}_{\dot{\alpha}} = e_i \wedge (-i \overline{Q}_{\dot{\alpha}} - i \overline{\sigma}^{\mu}_{\dot{\alpha} \beta} \theta^{\beta} \partial_{\mu} ) = i e_i \wedge \overline{Q}_{\dot{\alpha}} - i \overline{\sigma}^{\mu}_{\dot{\alpha}\beta}z^{\beta j} e_i \wedge e_j \partial_{\mu} =  \label{DotDisplacementProof} \end{equation}  
\begin{equation} =  -i e_i \wedge \overline{Q}_{\dot{\alpha}} - i \overline{\sigma}^0_{\dot{\alpha} \beta} z^{\beta j} \bigg( - \frac{i}{4} [e_i \wedge Q_1, e_j \wedge \overline{Q}_{\dot{1}}]_{\wedge} - \frac{i}{4} [e_i \wedge Q_2, e_j \wedge \overline{Q}_{\dot{2}}]_{\wedge} \bigg)   \nonumber \end{equation}
\begin{equation} - i \overline{\sigma}^1_{\dot{\alpha} \beta} z^{\beta j} \bigg( - \frac{i}{4} [e_i \wedge Q_1, e_j \wedge \overline{Q}_{\dot{2}}]_{\wedge} - \frac{i}{4} [e_i \wedge Q_2, e_j \wedge \overline{Q}_{\dot{1}}]_{\wedge} \bigg)  \nonumber \end{equation}
\begin{equation} - i \overline{\sigma}^2_{\dot{\alpha} \beta} z^{\beta j} \bigg(\frac{1}{4} [e_i \wedge Q_1, e_j \wedge \overline{Q}_{\dot{2}}]_{\wedge} - \frac{1}{4} [e_i \wedge Q_2, e_j \wedge \overline{Q}_{\dot{1}}]_{\wedge} \bigg)  \nonumber \end{equation}
\begin{equation} - i \overline{\sigma}^3_{\dot{\alpha} \beta} z^{\beta j} \bigg( - \frac{i}{4} [e_i \wedge Q_1, e_j \wedge \overline{Q}_{\dot{1}}]_{\wedge} + \frac{i}{4} [e_i \wedge Q_2, e_j \wedge \overline{Q}_{\dot{2}}]_{\wedge}  \bigg) \nonumber \end{equation}
Thus, all of $ g + ((V_D)_{\mathbb{C}})^4 + \Big( \bigwedge^2 (V_D) \Big)_{\mathbb{C}} \otimes_{\mathbb{C}} (T \mathbb{R}^4)_{\mathbb{C}}$ may be reached. 
\end{proof}

\begin{thm} The general supersymmetric transformation takes the form $\theta' = \theta + \xi$, $\overline{\theta}^{\prime} = \overline{\theta} + \overline{\xi}$, $x^{\prime} = x + \eta^{\alpha} \wedge \sigma^{\mu}_{\alpha \dot{\beta}} \overline{\theta}^{\dot{\beta}} + \theta^{\beta} \wedge \sigma^{\mu}_{\beta \dot{\alpha}} \overline{\eta}^{\dot{\alpha}} + c$ where $c \in \Big( \bigwedge^2 (V_D) \Big)_{\mathbb{C}} \otimes_{\mathbb{C}} (T \mathbb{R}^4)_{\mathbb{C}}$.
\end{thm}
\begin{proof}The set of transformations described in the statement of the theorem is closed under $Q$ and $\overline{Q}$. This immediately implies that the set generated by operators of the form $\eta^{\alpha} \wedge Q_{\alpha}$ and $\overline{\xi}^{\dot{\alpha}} \wedge \overline{Q}_{\dot{\alpha}}$ is a subset of the full set. We shall now show that this subset is in fact equal to the full set. We have already proven in the previous theorem that the translations by $c \in V_D \wedge V_D$ are generated by those transformations. It remains to show that the ``rotation" terms $ \eta^{\alpha} \wedge \sigma^{\mu}_{\alpha \dot{\beta}} \overline{\theta}^{\dot{\beta}}$ and $\theta^{\beta} \wedge \sigma^{\mu}_{\beta \dot{\alpha}} \overline{\eta}^{\dot{\alpha}}$ are generated by them as well. We will show this by the following calculation
\begin{equation} \eta^{\alpha} \wedge \sigma^{\mu}_{\alpha \dot{\beta}} \overline{\theta}^{\dot{\beta}} \partial_{\mu} = \eta^{\alpha} \wedge Q_{\alpha} - i \eta^{\alpha} \wedge \partial_{\alpha} = \eta^{\alpha} \wedge Q_{\alpha} - i z_i^{\alpha} e_i \wedge \partial_{\alpha} = \end{equation}
\begin{equation} = \eta^{\alpha} \wedge Q_{\alpha} - i z_i^{\alpha} \bigg(  i e_i \wedge Q_{\alpha} - i \sigma^0_{\alpha \dot{\beta}} \overline{z}^{\dot{\beta} j} \bigg( - \frac{i}{4} [e_i \wedge Q_1, e_j \wedge \overline{Q}_{\dot{1}}]_{\wedge} - \frac{i}{4} [e_i \wedge Q_2, e_j \wedge \overline{Q}_{\dot{2}}]_{\wedge} \bigg)   \nonumber \end{equation}
\begin{equation} - i \sigma^1_{\alpha \dot{\beta}} \overline{z}^{\dot{\beta} j} \bigg( - \frac{i}{4} [e_i \wedge Q_1, e_j \wedge \overline{Q}_{\dot{2}}]_{\wedge} - \frac{i}{4} [e_i \wedge Q_2, e_j \wedge \overline{Q}_{\dot{1}}]_{\wedge} \bigg)  \nonumber \end{equation}
\begin{equation} - i \sigma^2_{\alpha \dot{\beta}} \overline{z}^{\dot{\beta} j} \bigg(\frac{1}{4} [e_i \wedge Q_1, e_j \wedge \overline{Q}_{\dot{2}}]_{\wedge} - \frac{1}{4} [e_i \wedge Q_2, e_j \wedge \overline{Q}_{\dot{1}}]_{\wedge} \bigg)  \nonumber \end{equation}
\begin{equation} - i \sigma^3_{\alpha \dot{\beta}} \overline{z}^{\dot{\beta} j} \bigg( - \frac{i}{4} [e_i \wedge Q_1, e_j \wedge \overline{Q}_{\dot{1}}]_{\wedge} + \frac{i}{4} [e_i \wedge Q_2, e_j \wedge \overline{Q}_{\dot{2}}] _{\wedge} \bigg) \nonumber \bigg) \end{equation}

\begin{equation} \theta^{\beta} \wedge \sigma^{\mu}_{\beta \dot{\alpha}} \overline{\eta}^{\dot{\alpha}}  \partial_{\mu} = i \overline{Q}_{\dot{\alpha}} - i \overline{\eta} ^{\dot{\alpha}} \wedge \overline{\partial}_{\dot{\alpha}} = \end{equation}
\begin{equation} =  i \overline{Q}_{\dot{\alpha}} - i \overline{z}_i^{\dot{\alpha}} \bigg(-i e_i \wedge Q_{\alpha} - i \sigma^0_{\alpha \dot{\beta}} z^{\beta j} \bigg( - \frac{i}{4} [e_i \wedge Q_1, e_j \wedge \overline{Q}_{\dot{1}}]_{\wedge} - \frac{i}{4} [e_i \wedge Q_2, e_j \wedge \overline{Q}_{\dot{2}}]_{\wedge} \bigg)  \nonumber \end{equation}
\begin{equation} - i \sigma^1_{\alpha \dot{\beta}} z^{\beta j} \bigg( - \frac{i}{4} [e_i \wedge Q_1, e_j \wedge \overline{Q}_{\dot{2}}]_{\wedge} - \frac{i}{4} [e_i \wedge Q_2, e_j \wedge \overline{Q}_{\dot{1}}]_{\wedge} \bigg)  \nonumber \end{equation}
\begin{equation} - i \sigma^2_{\alpha \dot{\beta}} z^{\beta j} \bigg(\frac{1}{4} [e_i \wedge Q_1, e_j \wedge \overline{Q}_{\dot{2}}]_{\wedge} - \frac{1}{4} [e_i \wedge Q_2, e_j \wedge \overline{Q}_{\dot{1}}]_{\wedge} \bigg)  \nonumber \end{equation}
\begin{equation} - i \sigma^3_{\alpha \dot{\beta}} z^{\beta j} \bigg( - \frac{i}{4} [e_i \wedge Q_1, e_j \wedge \overline{Q}_{\dot{1}}]_{\wedge} + \frac{i}{4} [e_i \wedge Q_2, e_j \wedge \overline{Q}_{\dot{2}}]_{\wedge}  \bigg) \bigg) \nonumber \end{equation} 
where the long equations were produced by the substitutions of Eq \ref{DisplacementProof} and \ref{DotDisplacementProof}. 
\end{proof}

\subsection{Integration over a superspace} \label{SUSYVolume}

Before we proceed, we have to agree on the product signs we will be using. We know from Theorem \ref{EitherProductTheorem} that
\begin{equation} \int (d \theta_1 \wedge \ldots \wedge  d \theta_n)* f(\theta_1, \ldots, \theta_n) = \int d \theta_1 * \ldots *d  \theta_n* f(\theta_1, \ldots, \theta_n)   \label{WedgeStarIntegral} \end{equation}
We recall from Section \ref{ProductSigns} that we agreed to use $\theta_1 \theta_2$ for the cases when the product can be either $\wedge$ or $*$. Therefore, we can write the above integral as 
\begin{equation} \int (d \theta_1 \ldots  d \theta_n)* f(\theta_1, \ldots, \theta_n) \end{equation}
However, despite the fact that 
\begin{equation} \epsilon_{\alpha \beta} \xi^{\alpha} * \eta^{\beta} =  \epsilon_{\alpha \beta} \xi^{\alpha} \wedge \eta^{\beta} \label{EpsilonStarWedge}\end{equation}
we can not equate $\xi * \eta$ with $\xi \wedge \eta$ since $\xi * \eta$ is defined as $h_{\alpha \beta} \xi^{\alpha}* \eta^{\beta}$ as opposed to $\epsilon_{\alpha \beta} \xi^{\alpha} * \eta^{\beta}$: 
\begin{equation} \xi * \eta =  h_{\alpha \beta} \xi^{\alpha} * \eta^{\beta}  \neq \epsilon_{\alpha \beta} \xi^{\alpha} * \eta^{\beta} = \xi \wedge \eta \end{equation}
where $h_{\alpha \beta}$ is a symmetric tensor as opposed to $\epsilon_{\alpha \beta}$ which is antisymmetric:
\begin{equation} h_{11}=h^{11}=h_{22}=h^{22}= 1 \end{equation}
\begin{equation} \overline{h}_{\dot{1} \dot{1}}= \overline{h}^{\dot{1} \dot{1}} = \overline{h}_{\dot{2} \dot{2}} = \overline{h}^{\dot{2} \dot{2}} = 1 \end{equation}
\begin{equation} h_{12} = h_{21} = h^{12} = h^{21} = 0 \end{equation}
\begin{equation} \overline{h}_{\dot{1} \dot{2}} = \overline{h}_{\dot{2} \dot{1}} = \overline{h}^{\dot{1} \dot{2}} = \overline{h}^{\dot{2} \dot{1}} = 0 \end{equation}
\begin{equation} \epsilon^{12} = - \epsilon^{21} = \epsilon_{21} = - \epsilon_{12} = 1 \end{equation}
\begin{equation} \overline{\epsilon}^{\dot{1} \dot{2}} = - \overline{\epsilon}^{\dot{2} \dot{1}} = \overline{\epsilon}_{\dot{2} \dot{1}} = - \overline{\epsilon}_{\dot{1} \dot{2}} = 1 \end{equation}
\begin{equation} \epsilon^{11} = \epsilon^{22} = \epsilon_{11} = \epsilon_{22} = 0 \end{equation} 
\begin{equation} \overline{\epsilon}^{\dot{1} \dot{1}} = \overline{\epsilon}^{\dot{2} \dot{2}} = \overline{\epsilon}_{\dot{1} \dot{1}} = \overline{\epsilon}_{\dot{2} \dot{2}} = 0 \end{equation} 
 Those definitions do not violate SU(2) symmetry since the $\sigma$-matrices that generate it, $\sigma^{\mu}_{\alpha \dot{\beta}}$ and $\overline{\sigma}^{\mu}_{\dot{\alpha} \beta}$, include both dotted and undotted indexes, whereas the $\epsilon$-tensors and $h$-tensors include either only dotted indexes or only undotted ones. We can streamline the notation by agreeing that when $\eta^2$ is written, $\epsilon_{\alpha \beta}$ is meant, in contrast to $\vert \eta \vert^2$ that involves $h_{\alpha \beta}$: 
\begin{equation} \eta^2 = \epsilon_{\alpha \beta} \eta^{\alpha} \eta^{\beta} \; , \; \vert \eta \vert^2 = h_{\alpha \beta} \eta^{\alpha} * \eta^{\beta} \end{equation}
where the product sign is skipped at the left in light of Eq \ref{EpsilonStarWedge}. To define a superfield, we follow the usual steps and write 
\begin{equation} \Phi (x_L, \theta) = \phi (x_L) + \sqrt{2} \theta^{\alpha} \wedge \psi_{\alpha} (x_L) + \theta^2 F (x_L) \label{Phi} \end{equation}
where 
\begin{equation} x_L^{\mu} = x^{\mu} - i \theta^{\alpha} \wedge \sigma^{\mu}_{\alpha \dot{\beta}} \overline{\theta}^{\dot{\beta}} \; , \; x_R^{\mu} = x^{\mu} + i \theta^{\alpha} \wedge \sigma^{\mu}_{\alpha \dot{\beta}} \overline{\theta}^{\dot{\alpha}} \end{equation}
In the same way as it is done conventionally, we are assuming that the functions in Eq \ref{Phi} are analytic extensions from usual space to superspace. This would dictate to us that, for example, 
\begin{equation} \phi (x_L) = \phi (x) -  i \theta^{\alpha} \sigma^{\mu}_{\alpha \dot{\beta}} \overline{\theta}^{\dot{\beta}} \partial_{\mu} \phi \label{AnalyticExtension} \end{equation}

However, in light of more realistic model of Grassmann numbers that we introduce here, the functions do not a priori have to be analytic (see Chapter \ref{ChapterNonAnalytic}). At the same time, we still need to assume analyticity in the integration below, in order to recover the non-supersymmetric Lagrangians. What this means is that analyticity will become a \emph{constraint} on the path integral, as opposed to a mathematical fact. In other words, we will \emph{not} be integrating over all field trajectories over a superspace; but, instead, we will only integrating over \emph{analytic} ones. However, this only requires analyticity over $\theta$ and $\overline{\theta}$, but it does not require analyticity over $x$. On the other hand, this requires the derivative over $x$ to be well defined, since the above formulae includes it. This, in itself is a problem: in order for our description of measure over Grassmann numbers to be applicable, we need the fermionic fields at the nearby spacetime points to be completely independent of each other, which contradicts the assumption of analyticity. 

One way to handle this is to perform a Fourier transform, impose our Grassmann measure on a Fourier space, and impose an upper bounds on frequency. Another option is to discretize space, in which case the derivatives over $x$ would be defined as differences. If we make the second choice, then our superspace would consist of countably many parallel hyperplanes. Each hyperplane would project onto a single lattice point in $\mathbb{R}^4$. Those hyperplanes would be continuous and, therefore, the supersymmetry transformations (which do not affect rank 0 components but only affect ranks 1 and 2) would map each hyperplane onto itself. The function will be called analytic if it obeys a \emph{discretized} version of Eq \ref{AnalyticExtension}: that is, we replace all the derivatives in Eq \ref{AnalyticExtension} with corresponding differences. The simplest way of doing so is to write 
\begin{equation} \phi (x_L) = \phi (x) -  i \theta^{\alpha} \sum_{\mu=0}^3 \sigma^{\mu}_{\alpha \dot{\beta}} \overline{\theta}^{\dot{\beta}} \frac{\phi (x+ \epsilon \hat{x}_{\mu}) - \phi (x)}{\epsilon} \label{AnalyticExtension} \end{equation}
where $\hat{x}_{\mu}$ is a unit vector in $\mu$-direction and $\epsilon$ is some scaling size. Since this clearly violates rotational invariance, there is a more rotationally invariant alternative to this. In particular, we can perform the following calculation:
\begin{equation} \int_{- \infty}^{\infty} (x'-x) f(x') e^{- \frac{\alpha}{2} (x'-x)^2} dx' = \int_{- \infty}^{\infty} (x'-x)^2 f^{\prime} (x) e^{- \frac{\alpha}{2} (x'-x)^2} d(x'-x)  \nonumber \end{equation} 
\begin{equation} = f^{\prime} (x) \int_{- \infty}^{\infty} (x'-x) e^{- \frac{\alpha}{2} (x'-x)^2} d \frac{(x'-x)^2}{2} = - \frac{f^{\prime} (x)}{\alpha} \int_{- \infty}^{\infty} (x'-x) d e^{- \frac{\alpha}{2} (x'-x)^2}   \nonumber \end{equation} 
\begin{equation} = \frac{f^{\prime} (x)}{\alpha} \int_{- \infty}^{\infty} e^{- \frac{\alpha}{2} (x'-x)^2} dx = \sqrt{\frac{2 \pi}{\alpha}} \frac{f^{\prime} (x)}{\alpha} \end{equation}
This implies that 
\begin{equation} f^{\prime} (x) = \frac{\alpha^{3/2}}{\sqrt{2 \pi}} \int_{- \infty}^{\infty} (x'-x) f(x') e^{- \frac{\alpha}{2} (x'-x)^2} dx \end{equation}  
On the other hand, we can write
\begin{equation} f(x) = \sqrt{\frac{\alpha}{2 \pi}} \int_{- \infty}^{\infty} e^{- \frac{\alpha}{2} (x-x')^2} dx \end{equation}
Therefore, 
\begin{equation} \partial_{\mu} f = \frac{1}{f^3 (x) } (\partial_{\mu} f) \prod_{\nu \neq \mu} f  \nonumber \end{equation}
\begin{equation} =  \frac{1}{f^3 (x)} \bigg(\frac{\alpha^{3/2}}{\sqrt{2 \pi}} \int_{- \infty}^{\infty} f(x') (x^{\prime \mu}-x^{\mu}) e^{- \frac{\alpha}{2} (x^{ \prime \mu}-x^{\mu})^2} dx \bigg) \prod_{\nu \neq \mu} \bigg( \sqrt{\frac{\alpha}{2 \pi}} \int_{- \infty}^{\infty} e^{- \frac{\alpha}{2} (x-x')^2} dx \bigg)  \nonumber \end{equation} 
\begin{equation} = \frac{1}{f^3 (x)} \frac{\alpha^3}{4 \pi^2} \int f(x') (x^{ \prime \mu} - x^{\mu}) e^{- \frac{\alpha}{2} \vert \vec{x}^{\prime} - \vec{x} \vert^2} d^4 x^{\prime} \end{equation} 
We can now replace integral with a sum and write 
\begin{equation} \partial_{\mu} f = \frac{1}{f^3 (x)} \frac{\alpha^3}{4 \pi^2} \sum f(x') (x^{\prime \mu} - x^{\mu}) e^{- \frac{\alpha}{2} \vert \vec{x}^{\prime} - \vec{x} \vert^2} \delta v \end{equation} 
where $\delta v$ is an element of four-volume taken up by a single point. Thus, the constraint given in Eq \ref{AnalyticExtension} becomes
\begin{equation} \phi (x_L) = \phi (x) -  i \theta^{\alpha} \sigma^{\mu}_{\alpha \dot{\beta}} \overline{\theta}^{\dot{\beta}}  \frac{1}{f^3 (x)} \frac{\alpha^3}{4 \pi^2} \sum f(x') (x^{\prime \mu} - x^{\mu}) e^{- \frac{\alpha}{2} \vert \vec{x}^{\prime} - \vec{x} \vert^2} \delta v \label{AnalyticExtensionDiscrete} \end{equation}
This would still violate relativity. After all, the Lorentzian $\epsilon$-neighborhood is the vicinity of the light cone, given by its ``past" component, 
\begin{equation} -   \sqrt{ \vert \vec{x} \vert^2 + \epsilon^2} \leq t \leq - \vert \vec{x} \vert \end{equation}
and its ``future" component, 
\begin{equation} \vert \vec{x} \vert \leq t \leq \sqrt{ \vert \vec{x} \vert^2 + \epsilon^2} \end{equation} 
both of which can be shown to have infinite volume which implies that each point would have infinitely many neighbors. Therefore, any method of making the number of neighbors finite would violate relativity by default. This, indeed, is a widely acknowledged problem among people working with discrete spacetimes (see \cite{Nonlocality}). However, for our purposes, we are content with this. 

In light of discreteness of spacetime, we replace the integral over $x$ with a sum over $x$ in our supersymmetric action, while the integrals over $\theta$ and $\overline{\theta}$ will remain integrals (since $\theta$ and $\overline{\theta}$ remain continuous). Thus, 
\begin{equation} S =  \sum_x (\delta v)_x \int (d^2 \overline{\theta} d^2 \theta)*( \Phi (x, \theta, \overline{\theta}) \wedge \overline{\Phi} (x, \theta, \overline{\theta}))  \nonumber \end{equation}
\begin{equation} +  \sum_x (\delta v)_x  \int d^2 \theta * ({\mathcal W} (\Phi))(x, \theta ) + \sum_x (\delta v)_x \int  d^2 \overline{\theta} * (\overline{\mathcal W} (\overline{\Phi}))(x,  \overline{\theta}) \label{DifferentVolumeDimensions} \end{equation}
where ${\mathcal W} (\Phi)$ and $\overline{\mathcal{W}} (\overline{\Phi})$ are both expressed in terms of wedge-products:
\begin{equation} {\mathcal W} (\Phi) = \frac{m}{2} \Phi \wedge \Phi - \frac{\lambda}{3} \Phi \wedge \Phi \wedge \Phi \end{equation} 
\begin{equation} \overline{{\mathcal W}}  (\overline{\Phi}) = \frac{m}{2} \overline{\Phi} \wedge \overline{\Phi} - \frac{\lambda}{3} \overline{\Phi} \wedge \overline{\Phi} \wedge \overline{\Phi} \end{equation} 
and the differentials are defined as 
\begin{equation} d^2 \theta = \epsilon_{\alpha \beta} d_{\mu^*} \theta^{\alpha} d_{\mu^*} \theta^{\beta} \; , \; d^2 \overline{\theta} = \overline{\epsilon}_{\dot{\alpha} \dot{\beta}} d_{\mu^*} \overline{\theta}^{\dot{\alpha}} d_{\mu^*} \overline{\theta}^{\dot{\beta}} \end{equation}
To make it even more explicit, we substitute Eq \ref{DirectedMeasure}, 
\begin{equation} d_{\mu} \theta = k_D \theta e^{- \alpha \vert \theta \vert^2/2} dV  \end{equation}
to obtain
\begin{equation} d_{\mu} \theta^2 = \epsilon_{\alpha \beta} d_{\mu} \theta^{\alpha} d_{\mu} \theta^{\beta}= k_D^2 \epsilon_{\alpha \beta} \theta^{\alpha} \theta^{\beta} e^{- \alpha \vert \theta \vert^2/2} dV \end{equation}
where we have $e^{- \alpha \vert \theta \vert^2/2}$ instead of $e^{- \alpha \vert \theta \vert^2}$ and $dV$ instead of $(dV)^2$ because we are combining $D$-dimensional $\theta_1$ with $D$-dimensional $\theta_2$ into $2D$-dimensional $\theta$. To emphasize this point, we will write $d^{2D} V$. Similarly, 
\begin{equation} d_{\mu} \overline{\theta}^2 = k_D^2 \overline{\epsilon}_{\dot{\alpha} \dot{\beta}} \theta^{\dot{\alpha}} \theta^{\dot{\beta}} e^{- \alpha \vert \overline{\theta} \vert^2/2} dV \end{equation} 
By substituting those, we obtain 
\begin{equation} S =  k_D^4 \sum_x (\delta v)_x \int d^{4D} V e^{- \alpha \vert \theta \vert^2/2 - \alpha \vert \overline{\theta} \vert^2/2} (\epsilon_{\alpha \beta} \epsilon_{\dot{\gamma} \dot{\delta}} \theta^{\alpha} \theta^{\beta} \overline{\theta}^{\dot{\gamma}} \overline{\theta}^{\dot{\delta}})*( \Phi (x, \theta, \overline{\theta}) \wedge \overline{\Phi} (x, \theta, \overline{\theta}))  \nonumber \end{equation}
\begin{equation} +  k_D^2 \sum_x (\delta v)_x  \int d^{2D} V  e^{- \alpha \vert \theta \vert^2/2} (\epsilon_{\alpha \beta} \theta^{\alpha} \theta^{\beta})* ({\mathcal W} (\Phi))(x, \theta )  \nonumber \end{equation}
\begin{equation} + k_D^2 \sum_x (\delta v)_x \int  d^{2D} V e^{- \alpha \vert \overline{\theta} \vert^2/2} (\overline{\epsilon}_{\dot{\alpha} \dot{\beta}} \overline{\theta}^{\dot{\alpha}} \overline{\theta}^{\dot{\beta}})* (\overline{\mathcal W} (\overline{\Phi}))(x,  \overline{\theta}) \label{DifferentVolumeDimensions} \end{equation}
On the other hand, if we are to use the surface integral instead of volume integral, then $D$ gets replaced by $D-1$, and, accordingly, $2D$ and $4D$ gets replaced by $2D-2$ and $4D-4$, respectively, which represents the dimensionality of tori. The coefficients $k_D$ along with the factors $e^{- \alpha \vert \theta \vert^2/2}$ get dropped. So we obtain 
\begin{equation} S =  \frac{1}{V^4} \sum_x (\delta v)_x \int d^{4D-4} V  (\epsilon_{\alpha \beta} \overline{\epsilon}_{\dot{\gamma} \dot{\delta}} \theta^{\alpha} \theta^{\beta} \overline{\theta}^{\dot{\gamma}} \overline{\theta}^{\dot{\delta}})*( \Phi (x, \theta, \overline{\theta}) \wedge \overline{\Phi} (x, \theta, \overline{\theta})) \nonumber \end{equation}
\begin{equation} +  \frac{1}{V^2} \sum_x (\delta v)_x  \int d^{2D-2} V  (\epsilon_{\alpha \beta} \theta^{\alpha} \theta^{\beta})* ({\mathcal W} (\Phi))(x, \theta ) \nonumber \end{equation}
\begin{equation} + \frac{1}{V^2} \sum_x (\delta v)_x \int  d^{2D-2} V  (\overline{\epsilon}_{\dot{\alpha} \dot{\beta}} \overline{\theta}^{\dot{\alpha}} \overline{\theta}^{\dot{\beta}})* (\overline{\mathcal W} (\overline{\Phi}))(x,  \overline{\theta}) \label{DifferentSurfaceDimensions} \end{equation}
Thus, in case of the volume integral we have discrete copies of hyperplanes and in case of the surface integral we have discrete copies of tori. It is important to note that the torus does not map onto itself under supersymmetry transformation. At the same time, the statement that the Lagrangian ``happens" to be invariant under the supersymmetry transformation continues to be true: it is simply irrelevant to the integral we are taking. 

The other possibility that we briefly talked about is the minimalist model (section \ref{Misc1}): namely that $V_D$ is viewed as $\{1/2, -1/2 \}^D$. If we take this approach then the supersymmetric will simply add either $1/2 e_k$ or $-1/2 e_k$ to each rank-1 components and, therefore, the rank 1 will be half-inegers. As far as rank-2 components, they would be changed by the products of rank-1 components, which will be half-integers and, therefore, they will be quarter-integers. Despite the fact that rank-1 components will take all half-integer values, the ``integral" over the superspace will only sum their values of $+1/2$ and $-1/2$. This is reminiscent to what we said for the case of the surface integral where, despite the fact that suersymmetric transformation takes us away from the surface, we are only integrating over the surface.  

As far as the integral over $x$ goes, the inspection of the equation
\begin{equation} \int (d^4 x d^2 \overline{\theta} d^2 \theta) * (\overline{\Phi} \wedge \Phi )= \int d^4 x (\partial_{\mu} \overline{\phi} \partial^{\mu} \phi + i \overline{\psi}^{\dot{\alpha}} \wedge \overline{\sigma}^{\mu}_{\dot{\alpha} \beta} \psi^{\beta} + \overline{F} \wedge F) \end{equation}
tells us that since on the right hand side the integral over $x$ does not include rank 2 component, neither should the integral on the left hand side include a rank 2 component.  The way we interpret it is that -- in the case we chose to do volume integral -- we are taking the above integral over the hyperplane where the rank 2 component is equal to zero. In the case we chose to do surface integral, we are integrating over a ``hypercircle" within this hyperplane -- instead of integrating over a hypersphere inside the whole space; and then we produce the torus as a product of these hypercircles. In terms of discretization of $x$, this means that, instead of integrating over the hyperplanes (hyperspheres) given above, we will be integrating over their subspaces (subsets), where rank-1 components can alter but rank-2 components can not. The problem with doing that is that it violates supersymmetry. In particular, since supersymmetric transformations alter rank 2 components, the hyperplane (hypercircle) defined by rank 2 components being equal to zero would \emph{not} map onto itself. But that is where our analyticity constraint comes to the rescue. It can be shown that \emph{as long as the function is analytic}, its integral over the above hyperplane (hypercircle) would be the same as its integral over its supersymmetric transformation. Thus, it does not matter which of those hyperplanes (hypercircles) we would select: we would get the same answer!

 Indeed, one has to get used to the idea of selecting hyperplanes (hypercircles) inside our space (sphere), since one has to do it in order to make sense why the last two terms of Eq \ref{DifferentVolumeDimensions} have fewer differentials than the first term on the right hand side. That feature might be bothersome in its own right. What are we physically saying? Are we saying that there is some special hyperplane (hypercircle) that plays some physical role? The answer is no. In light of the fact that we have introduced the symmetries, the value of the integral is independant of the choice of that hyperplane (hypercircle). This implies that if we were to integrate it over the whole space (sphere), we would have infinite overcounting over infinitely many copies of the same integral, leading to an infinite answer. In order to avoid the infinite overcounting, we are restricting that integral to a choice of hyperplane (hypercircle) -- and it does not matter which choice we make. 

As far as discretization of $x$, it makes one wonder why are not we equally explicit with regards to discretizing $\theta$. After all, from a mathematically rigorous point of view, the discretization is a necessary component of a path integral. The answer to this question is two-fold. First of all, the reason for discretizing $x$ is different than the one for discretizing $\theta$. In particular, we want $x$ discretized so that the $\theta$-s at the two neighboring $x$-points can be, independently, integrated over the whole sphere (if we choose the option of the surface integral) or that the non-trivial formula for measure of the integration at any point is not influenced by its neighbor (if we choose the option of the volume integral). On the other hand, discretizing $\theta$ has to do with discretizing either the sphere or the volume we just mentioned. From this point of view, it is perfectly reasonable to talk about the degree of approximation that overlooks discretization of $\theta$ without overlooking discretization of $x$. Indeed, doing so is paedagogical since it emphasizes the point we are making. The other answer to this question is that -- thanks to the constraints we talked about -- $\theta$ and $\overline{\theta}$ do not have to be discretized at all! Even if the spheres or hyperplanes on which $\theta$ and $\overline{\theta}$ reside have uncountably many points, we only need a very small number of parameters to completely characterize the function over any given sphere. So if we take the integral over those parameters, that integral will be well defined \emph{even in the case of the sphere being continuous}, by the virtue of the fact that the number of those parameters remains finite. Of course, this statement would nor apply to the non-analytic cases (such as discussed in the next chapter). But the constraint we described earlier, if left unmodified, implies analyticity. So -- again if we leave everything unmodified -- we would be able to make the argument we just did in favor of continuous sphere (hyperplane). If we do, however, choose to modify things and introduce non-analytic functions, then the sphere (hyperplane) would have to be discretized as well. 

\section{Extension of superanalysis to non-analytic functions} \label{ChapterNonAnalytic} 

\subsection{Examples of non-analytic functions} \label{ExamplesNonlinear}

The Riemann sum definition of the Berezin integral implies that it makes sense to integrate both analytic and non-analytic functions. Since the specific choice of a set of measures (or choice of set closed surfaces) was designed for analytic functions, its not a big surprise that it might not work as nicely for non-analytic ones. To be sure, if one describes a specific surface, both analytic and non-analytic functions can be integrated over it. But if, on the other hand, the only information that is given is that the surface encloses a volume of $1/D$, then it is possible that different surfaces that meet that description would result in vastly different values an integral of non-analytic function can take. Nevertheless, there are some types of non-analytic functions for which we can get well defined integrals. Let us discuss some of them. 

One example is if we use Clifford product, instead of wedge product, in Taylor series. In this case, we can use Equation \ref{avgint} to show that the integral would approach a linear coefficient in the limit of $D \rightarrow \infty$. However, the function does not have to be analytic with respect to Clifford product either. A radically different example of a function that can be integrated is 
\begin{equation} f \bigg( \sum_{k \in \mathbb{N}} x_k e_k \bigg) = \sum_{k \in S} x_k e_k  \end{equation}
for some $S \subset \mathbb{N}$. In this case one can show that, provided the limit on the right hand side below exists, the integral evaluates to 
\begin{equation} \lim_{D \rightarrow \infty} \int d_{\mu_D} \theta \; f (\theta) = \lim_{D \rightarrow \infty} \bigg(\frac{1}{D} \sharp \{k \in S, k <D \} \bigg) \end{equation}
where $\sharp$ stands for the number of elements. One could also extend our definition of integral to include Lebesgue integrals, and define $f (\theta)$ in such a way that $f(x_1 e_1 + \cdots + x_D e_D)$ is $1$ if all values of $x_k$ are rational and $0$ otherwise. One can show that the integral of this function is zero. Obviously, the list of a few non-analytic functions we just provided is by no means exclusive but, hopefully, it can convince the reader that one can obtain well defined results with those functions. 

\subsection{Physics applications of non-analytic functions} \label{PhysicsNonlinear} 

The second author is undertaking companion projects where non-analytic functions are applied to physics. One application is the theory of quantum measurement. According to the deBroglie's \cite{DeBroglie} Bohm's \cite{Bohm1, Bohm2} interpretation of quantum mechanics, a particle and a wave exist at the same time as a separate entities, and the wave guides a particle through an expression of the form $dx/ dt = - \nabla S/ m$. If by $\rho$ one means the classical probability of finding the particle, and $\psi$ is a guiding wave, then $\rho = \vert \psi \vert^2$ was shown to arize as an equilibrium probability distribution. This approach was then generalized to fields by replacing $\psi (x)$ with $\psi (\phi)$ and $dx/dt$ with $d \psi/ dt$ \cite{Struyve,Holland1, Holland2, Valentini} Since fermionic fields are Grassmann-valued, most approaches evade them. For example, Bell \cite{Bell} modeled bosons as fields and fermions as particle numbers at the lattice points. Struyve and Westmann \cite{Struyve} on the other hand proposed a model with bosons being the only observables (for example, we do not observe an electron hitting the screen, we observe the photons it emits). Valentini, on the other hand, attempted to include fermionic feables (see Sec 4.2 of \cite{Valentini}) and he has shown that both a guidance equation and a continuity equation may be formally written down. However, those equations lose their physical meaning since, for one thing, Grassmann numbers can not take distinguishable values. Apart from that, the function $\rho (\psi)$ would have to be linear, which would force us to say that $\psi$ would take infinitely large values with infinite probability, in contradiction to anything we would expect. However, in light of the proposal presented in the current paper, there is a way to address those issues. We now say that a fermionic field takes well defined values that live in $D$-dimensional space we described. Furthermore, since we have a concept of non-analytic functions, we no longer have to assume that probability distribution is linear and, instead, we can use a continuity equation (that now regains its meaning) to investigate various non-linear probability distributions we might have. That might be especially useful if we attempt to couple those fields to gravity \cite{PilotWaveGravity}. 

Another set of approaches to quantum measurement that might also benefit from this paper are Ghirardi Rimini Weber (GRW) models  \cite{GRW1, GRW2} and weighted path integrals \cite{Mensky1, Mensky2, Kent, Epsilon}. The quantum mechanics version of GRW model \cite{GRW1, GRW2} is that we have a wave function that evolves according to Schrodinger equation and that evolution is being interrupted by so-called ``hits" when the wave function is being multiplied by Gaussians, and those Gaussians account for why classical objects remain on well defined locations instead of diffusing due to Schrodinger equation. This idea can be then extended to quantum field theory by replacing $\psi (x)$ with $\psi (\phi)$ \cite{GRWvsMensky}. On the other hand, the concept of weighted path integral \cite{Mensky1, Mensky2, Kent, Epsilon} is that, instead of having hits at random points in time, the measurement takes place continuously. Its outcome is a classical trajectory $\phi = \phi_{cl}$, and the probability of that trajectory is given by 
\begin{equation} \rho (\phi_{cl}) = \int [{\mathcal D} \phi] w (\phi, \phi_{cl}) e^{iS (\phi)} \end{equation}
where $w (\phi, \phi_{cl})$ is a \emph{weight function}, typically given as
\begin{equation} w (\phi, \phi_{cl}) = \exp \bigg( -\frac{\alpha}{2} \int (\phi - \phi_{cl})^2 d^4 x \bigg) \end{equation}
It has been shown that continuous measurement model can, in fact, arise out of GRW in a limit of very large time or very frequent hits \cite{GRWvsMensky}. Be it as it may, applying this to fermionic field raises a lot of questions. Whether we use hits or a weight function, in both cases those are Gaussians. However, Gaussians on anticommuting space are either constants or linear functions and, therefore, they lose their purpose. Nevertheless, within the context of our current paper, we have much better version of Gaussians: the ones that are based on Clifford product. This would, in turn, allow us to explore the fermionic version of those models. 

The other application is the causal set theory \cite{CausalSetReview1, CausalSetReview2, CausalSetReview3, CausalSetReview4}. A causal set is based on the observation that the metric of spacetime up to scaling can be inferred from lightcone causal relations \cite{Hawking,Malament}. Consequently, it was suggested to view spacetime as a partially ordered set, or \emph{causal set}, where the partial ordering is identified with the lightcone causal structure. One appealing feature of this is that partial ordering, as defined, is manifestly Lorentz invariant -- in sharp contrast with other discrete structures such as cubic lattices that have preferred directions. To keep in line with this, it is assumed that points are distributed randomly via Poisson process, which seems to be the most relativistic version of discrete spacetime one can find. This, however, creates locality problems. Lorentzian neighborhood is a vicinity of light cone. That, in turn, stretches arbitrarily far coordinate-wise and has infinite volume. If there was a preferred frame (such as the case in a cubic lattice) one could argue that the edges of the light cone -- despite small Lorentzian distance -- will not be connected to the origin. This can no longer be said if we have a truly random distribution of points. In \cite{BosonicCausalSet1, BosonicCausalSet2, BosonicCausalSet3,BosonicCausalSet4} it was suggested to use the actual field trajectory as a means of violating relativity and restoring locality. Thus, there is no afore-given preferred frame but, instead, preferred frame is a function of a specific field trajectory and, if we take path integral over all possible field trajectories, we would go over all possible preferred frames. However, this statement amounts to saying that the coefficients that are responsible for coupling of fields between the neighboring points are, themselves, functions of the field trajectory. This implies that the Lagrangian is no longer truly quadratic: the ``coefficients" in what ``looks like" a quadratic function are, themselves, functions! This, in turn, raises a question: how can a fermionic field be modeled this way? The only fermionic model that was proposed \cite{FermionicCausalSet} was a toy model where fermionic fields are commuting. However, in light of the paper at hand, one no longer needs to stick to a toy model and, instead, introduce actual fermionic fields on a causal set. This paper allows us to introduce a non-analytic function on anticommuting set and, in particular, we could search for the type of non-analytic Lagrangian for fermions that would accomplish the above goals.

One should not become too optimistic, however. As far as the Bohmian model goes, it does not propose any modifications to the wave function; it simply ``adds" a beable that the wave function carries. Its key idea is that wave function naturally splits into branches by the sole merit of its unitary evolution, and the particle will end up occupying one of those branches. However, that statement about branches is clearly not true when it comes to Grassmann variables, since analytic functions are linear in them. Of course, we could still have branches in other variables (such as bosonic field that interacts with fermionic field) but then we would be able to do ``just as much" as before: there is no clear way in which the interpretation of Grassmann variables really added anything. As far as GRW model is concerned, its key idea is that there is a modification to the unitary evolution of a wave function that would add up to produce a large effect. This seems to be harder to believe when it comes to Grassmann variables. After all, we need to retain the measure, which includes the fact that it is a function symmetric around the origin. On the other hand, if through some accumulation of small effects it would ``collapse" around some point away from the origin, that symmetry would be violated. If we could find a way to alter the direction of the measure as well, then we could make the measure symmetric around that new point -- and then the integrals would again coincide with Berezin since they would be shifted by a constant that integrates to zero. But then we are back with a question: how would that collapse be physically observed? We could say it is observed through some indirect means such as gravity, but then in what physical sense is it a fermionic field as opposed to some other hidden variable coupled to it? As far as the causal set idea is concerned, we were focusing on how to make a function smooth in some discretized notion of smoothness. But that would again contradict what we said about Grassmann measure: the measure on the space describing $\psi (x)$ would be concentrated around the average value of $\psi$ over neighboring points of $x$, and would no longer be centered around the origin. Nevertheless, one might attempt to identify physically observable parameters with small alterations to analyticity. This would probably amount to the fact that the above ideas would need to be replaced with subtler versions of themselves.

\section{Conclusion}

In this paper we have shown that the Berezin integral can, in fact, be represented as a limit of Riemann sums if we view it as a geometric integral over a closed surface that encloses a volume of $1/D$, taken in the limit as $D \rightarrow \infty$. If $D$ is finite, then the deviation between the geometric integral and the Berezin integral will be small if $D$ is much larger than the number of iterations of integral sign. 

However, the closed surface interpretation of the integral does not respect changes of variables. Thus, an alternative model is proposed where, instead of a surface integral, we have a volume integral, but the volume element carries a direction. The direction of a volume element plays the same role as a direction of a surface element which results in the integral having expected properties, with an added bonus that it respects change of variables. The price to pay for this, however, is that directed volume is not a usual occurrence in multivariable calculus the way directed area is.

One application is that the concept of supermanifolds can be interpreted as a usual continuous manifold. In particular, a supermanifold with $m$ commuting dimensions and $n$ anticommuting ones will be re-interpreted as an ordinary manifold with $m+nD$ dimensions. The integration over all of the coordinates of the manifold will be reinterpreted as an integral that has $m+n(D-1)$ dimensional hypersurface. Even though $m+n(D-1) \neq (m+nD)-1$ for $n \neq 1$, the normal vector will be well defined: in particular, it would be a sum of the normal vectors to $D-1$ dimensional hypersurfaces inside corresponding $D$ dimensional submanifolds. This kind of reinterpretation would be particularly helpful for students trying to study supermanifolds since they would look more geometrically similar to the manifolds they are used to. 

Another application of what we have done is that the definition of the Berezin integral has been extended to non-analytic functions. Since we have a coordinate system, we can write down non-analytic functions for each $D$. If we also specify our choice of a surface for each $D$, we can use geometric calculus to evaluate the integral, $I_D$, over that surface. If it happens that $I_D$ approaches a specific value as $D \rightarrow \infty$, we can think of that as the ultimate value of the integral we are looking for. 

As we noted above, the second author is presently investigating possible physical applications of such an integration theory for non-analytic functions. 

Last but not least, the representation of the Berezin integral as a limit of Riemann sums presents foundational interest, akin to the definition of real numbers via Dedekind cuts. 

{\bf Acknowledgements} This paper originates in the second author's preprints \cite{Sv2} and \cite{Sv1}. The first author is partially supported by NSF DMS- 1800492 The second author was partially supported by DMS-1554456 and DMS 9215024. The physical interpretation of this work in the present paper is due to the second author. The second author thanks the anonymous reviewers of an earlier version of this work for their suggestions which have been implemented in the present paper.

\bibliographystyle{plain}

\begin{thebibliography}{1}

 \bibitem{Rabin} Jeffrey M. Rabin `` The Berezin integral as a contour integral" Physica 15D (1985) 65 Conference: C83-12-14 EFI-84-4-CHICAGO

\bibitem{Book} Doran, Chris ``Geometric algebra for physicists" Cambridge : Cambridge University Press, 2003.

\bibitem{Drees}
Manuel~Drees \and Rohini Godbole~\and Probir~Roy.
\newblock {\em Theory and phenomenology of sparticles : an account of
  four-dimensional $N=1$ supersymmetry in high energy physics}.
\newblock World Scientific (Singapore), 2004.

\bibitem{Brom}
Alan Bromborsky.
\newblock An introduction to geometric algebra and calculus.
\newblock
  \url{http://www2.montgomerycollege.edu/departments/planet/planet/Numerical_Relativity/bookGA.pdf},
  2014.

\bibitem{Dine}
Michael Dine.
\newblock {\em Supersymmetry and string theory{:} Beyond the standard model}.
\newblock Cambridge University Press, Cambridge, 2007.

\bibitem{HS}
David Hestenes and Garret Sobczyk.
\newblock {\em Clifford algebra to geometric calculus}.
\newblock Fundamental Theories of Physics. D. Reidel Publishing Co., Dordrecht,
  1984.
\newblock A unified language for mathematics and physics.

\bibitem{LDG}
Anthony Lasenby, Chris Doran, and Stephen Gull.
\newblock Grassmann calculus, pseudoclassical mechanics, and geometric algebra.
\newblock {\em J. Math. Phys.}, 34(8):3683--3712, 1993.

\bibitem{PeSc}
Michael~E. Peskin and Daniel~V. Schroeder.
\newblock {\em An introduction to quantum field theory}.
\newblock Addison-Wesley Publishing Company, Advanced Book Program, Reading,
  MA, 1995.
\newblock Edited and with a foreword by David Pines.

\bibitem{Rogers}
Alice Rogers.
\newblock {\em Supermanifolds}.
\newblock World Scientific Publishing Co. Pte. Ltd., Hackensack, NJ, 2007.
\newblock Theory and applications.

\bibitem{Sv2}
Roman Sverdlov.
\newblock Use of sphere and curves to define {B}erezin integral.
\newblock \url{arXiv:0908.2605}, 2009.

\bibitem{Sv1}
Roman Sverdlov.
\newblock Realistic interpretation of {G}rassmann variables.
\newblock \url{arXiv:1202.4449}, 2012.

\bibitem{DeBroglie} de Broglie, L. (1927). ``La mécanique ondulatoire et la structure atomique de la matière et du rayonnement". Journal de Physique et le Radium. 8 (5): 225–241. Bibcode:1927JPhRa...8..225D. doi:10.1051/jphysrad:0192700805022500

\bibitem{Bohm1}  Bohm, D. (1952). ``A suggested Interpretation of the Quantum Theory in Terms of Hidden Variables, I". Physical Review. 85 (2): 166–179. Bibcode:1952PhRv...85..166B. doi:10.1103/PhysRev.85.166.

\bibitem{Bohm2}  Bohm, D. (1952). ``A suggested Interpretation of the Quantum Theory in Terms of Hidden Variables, II". Physical Review. 85 (2): 180–193. Bibcode:1952PhRv...85..180B. doi:10.1103/PhysRev.85.180.

\bibitem{Struyve} W. Struyve and H. Westman, “A minimalist pilot-wave model for quantum electrodynamics”, Proc. Roy. Soc. A 463, 3115-3129 (2007), and arXiv:0707.3487.

\bibitem{Holland1}  P.R. Holland, ``The Quantum Theory of Motion", Cambridge University Press,
Cambridge (1993)

\bibitem{Holland2}  P.R. Holland, Phys. Lett. A 128, 9 (1988)



\bibitem{Valentini} A. Valentini, ``On the Pilot-Wave Theory of Classical, Quantum and Subquantum
Physics", PhD. Thesis, International School for Advanced Studies, Trieste (1992),
online http://www.sissa.it/ap/PhD/Theses/valentini.pdf.

\bibitem{Bell} J.S. Bell, CERN preprint CERN-TH. 4035/84 (1984), reprinted in J.S. Bell, ``Speakable and unspeakable in quantum mechanics", Cambridge University Press, Cambridge, 173 (1987); Phys. Rep. 137, 49 (1986); in ``Quantum Implications", eds.
B.J. Hiley and F. David Peat, Routledge, London, p. 227 (1987); in ``John S. Bell
on The Foundations of Quantum Mechanics", eds. M. Bell, K. Gottfried and M.
Veltman, World Scientific, Singapore, 159 (2001)

\bibitem{GRW1} G.C. Ghirardi, A. Rimini and T. Weber, ``A model
for a unified quantum description of macroscopic
and microscopic systems", in Quantum Probability
and Applications, L. Accardi et al. (eds), Springer,
Berlin, 1985.

\bibitem{GRW2} G.C. Ghirardi, A. Rimini and T. Weber, ``Unified dynamics for microscopic and macroscopic systems", Phys. Rev. D 34, 470 (1986)

\bibitem{Mensky1}  M.B. Mensky, ``Quantum continuous measurements,
dynamical role of information and restricted path
integrals", in Proceedings TH2002 (International
Conference on Theoretical Physics) Supplement,
% Birkh¨auser 2003, and arXiv:quant-ph/0212112.

\bibitem{Mensky2} M.B. Mensky ``Quantum Measurement and Decoherence" Kluwer Academic Publishers 2000

\bibitem{Kent} A. Kent ``Path integrals and reality" arXiv:1305.6565.

\bibitem{Epsilon} R. Sverdlov ``Link between quantum measurement and the $i \epsilon$ term in the QFT propagator" Phys. Rev. D 90, 125020 (2014) and arXiv:1306.1948

\bibitem{GRWvsMensky} R. Sverdlov ``Connection between GRW "spontaneous collapse" and Mensky's "restricted path integral" models" Foundations of Physics, 46(7), 825-835 2016 and arXiv:1305.7516

\bibitem{PilotWaveGravity} R. Sverdlov``Can gravity be added to Pilot Wave models?" arXiv:1010.0580 

\bibitem{CausalSetReview1} J. Henson, The causal set approach to quantum gravity, arXiv:gr-qc/0601121

\bibitem{CausalSetReview2} D.D. Reid; Introduction to causal sets: an alternate view of spacetime structure; Canadian Journal of Physics 79, 1-16 (2001); arXiv:gr-qc/9909075

\bibitem{CausalSetReview3} R.D. Sorkin, Causal Sets: Discrete Gravity (Notes for the Valdivia Summer School), In Proceedings of the Valdivia Summer School, edited by A. Gomberoff and D. Marolf; arXiv:gr-qc/0309009

\bibitem{CausalSetReview4} L. Bombelli, J. Lee, D. Meyer, R.D. Sorkin, Spacetime as a causal set, Phys. Rev. Lett. 59:521-524 (1987)

\bibitem{Nonlocality} Rafael Sorkin ``Does locality fail at intermediate length-scales?" arXiv:grqc/0703099

\bibitem{Hawking} S W Hawking, A R King and P J McCarthy 1976 “A new topology for curved spacetime
which incorporates the causal, differential and conformal structures” J. Math. Phys. 17
174-181.

\bibitem{Malament} D Malament 1977 “The class of continuous timelike curves determines the topology of
spacetime” J. Math. Phys. 18 1399-1404.

\bibitem{BosonicCausalSet1} R. Sverdlov, L. Bombelli ``Gravity and Matter in Causal Set Theory" Class.Quant.Grav.26:075011,2009 and arXiv:0801.0240

\bibitem{BosonicCausalSet2} R.Sverdlov ``Dynamics for causal sets with matter fields: A Lagrangian-based approach" Contribution to the Proceedings of DICE2008, From Quantum Mechanics through Complexity to Space-time: The role of emergent dynamical structures (Castiglioncello, Italy, Sept. 2008) J.Phys.Conf.Ser.174:012019,2009 and arXiv:0905.1506

\bibitem{BosonicCausalSet3} R.Sverdlov ``Bosonic Fields in Causal Set Theory" arXiv:0807.4709

\bibitem{BosonicCausalSet4} R.Sverdlov ``Gauge Fields in Causal Set Theory" arXiv:0807.2066 

\bibitem{FermionicCausalSet} R.Sverdlov ``Spinor Fields in Causal Set Theory" arXiv:0808.2956

\bibitem{Vierbines} R.Sverdlov ``A Geometrical Description of Spinor Fields" arXiv:0802.1914

\bibitem{ContourSverdlov} R.Sverdlov ``Realistic interpretation of Grassmann variables" arXiv: 1202.4449

\bibitem{Feynmann} Feynman R, Leighton R, and Sands M. The Feynman Lectures on Physics . 3 volumes 1964, 1966. Library of Congress Catalog Card No. 63-20717

\bibitem{RefereeReference1} H. De Bie and F. Sommen, Spherical harmonics and integration in superspace, J. Phys. A
40, 7193-7212 (2007) [arXiv:0705.3148]

\bibitem{RefereeReference2} M. Kieburg, H. Kohler, and T. Guhr, Integration of Grassmann variables over invariant
functions on flat superspaces, J. Math. Phys. 50, 013528 (2009) [arXiv:0809.2674]

\bibitem{RefereeReference3} H. De Bie, D. Eelbode, and F. Sommen, Spherical harmonics and integration in superspace
II, J. Phys. A 42, 245204 (2009) [arXiv:0905.2092]

\bibitem{RefereeReference4} K. Coulembier and M. Kieburg, Pizzetti formulae for Stiefel manifolds and applications,
Letters in Mathematical Physics 105, 1333-1376 (2015) [arXiv:1409.8207]

\bibitem{BerezinBook}  F. A. Berezin, Introduction to Superanalysis, 1st edn (Dordrecht: D. Reidel Publishing Company)

\bibitem{NegativeDimension} G V Dunne,  I G Halliday  ``Negative Dimensional Integration II. Path integrals and fermionic equivalence" Physics Letters B, Volume 193, number 2,3 
(1987)

\bibitem{Statistical1} K. B. Efetov, Supersymmetry in Disorder and Chaos, Cambridge University Press,
Cambridge, 1st edition (1997)

\bibitem{Statistical2} G. Akemann, J. Baik, and P. Di Francesco (eds.), The Oxford Handbook of Random
Matrix Theory, (First Edition, Oxford University Press (2015)

\bibitem{ScaleRelativity} Laurent Notalle ``Scale Relativity and Fractal Spacetime", Imperial College Press, 2011

\bibitem{Version5} R. Sverdlov ``A use of geometric calculus to reduce Berezin integral to the limit of a Riemann sum" VERSION 5: arXiv:0908.2605v5 

\end{thebibliography}

\end{document}